\documentclass[preprint,prd,
amsmath,amssymb,amsthm,nofootinbib]{revtex4}
\usepackage[mathscr]{euscript}
\usepackage{bm}
\usepackage{textcomp}
\usepackage{graphicx}
\usepackage{subfigure}
\usepackage{overpic}
\usepackage{multirow}
\usepackage{floatrow}
\usepackage{float} 
\usepackage{amsmath,amssymb,amsthm}
\usepackage{cases}
\usepackage[colorlinks=true,linkcolor=red]{hyperref}
\newfloatcommand{capbtabbox}{table}[][\FBwidth]
\newcommand{\bea}{\begin{eqnarray}}
\newcommand{\eea}{\end{eqnarray}}
\newcommand{\beq}{\begin{equation}}
\newcommand{\eeq}{\end{equation}}

\def\/{\over}
\allowdisplaybreaks[4]
\usepackage{makecell}
\usepackage{rotating}
\begin{document}

\title{Power spectrum with $k^6$ growth for primordial black holes }
	
\author{Rongrong Zhai$^1$\footnote{rongrongzhai@foxmail.com},
	Hongwei Yu$^{1,2}$\footnote{Corresponding author: hwyu@hunnu.edu.cn}
 and  Puxun Wu$^{1,2}$\footnote{Corresponding author: pxwu@hunnu.edu.cn}
 }
\affiliation{$^1$Department of Physics and Synergetic Innovation Center for Quantum Effects and Applications, Hunan Normal University, Changsha, Hunan 410081, China\\
$^2$Institute of Interdisciplinary Studies, Hunan Normal University, Changsha, Hunan 410081, China
}

\begin{abstract}

The decrease of  both the rolling speed  of the inflaton and the sound speed of the curvature perturbations can amplify  the curvature perturbations during inflation so as to generate a sizable amount of primordial black holes.
In the ultraslow-roll inflation scenario, it has been found that the power spectrum of curvature perturbations has a $k^4$ growth.
In this paper, we  find that when the speed of sound decreases suddenly, the curvature perturbations becomes scale dependent in the infrared limit and  the power spectrum of the curvature perturbation only  has a $k^2$ growth.
 Furthermore, by studying  the evolution of  the power spectrum in the inflation model, in which   both the sound speed of the curvature perturbations and the rolling speed of the inflaton are reduced,
we find that the power spectrum is nearly scale invariant at the large scales  to satisfy the constraint from the cosmic microwave background radiation observations, and at the same time can be enhanced at the small scales  to result in an abundant formation of  primordial black holes. In the cases of the simultaneous changes of the sound speed and the slow-roll parameter $\eta$ and the change of the sound speed  preceding that of the slow-roll parameter $\eta$, the power spectrum can possess a  $k^6$ growth under certain conditions, which is the steepest growth of the power spectrum reported so far.

\end{abstract}

\maketitle
\section{Introduction}
\label{sec_in}

During the standard slow-roll inflation, the solution of  the Sasaki-Mukhanov equation for the evolution of  the curvature perturbations $\mathcal{R}$ contains, in the infrared limit, a constant term and a time-decaying one, and  this solution results  in a nearly scale-invariant power spectrum of   the curvature perturbations~\cite{Bruce A. Bassett,Antonio Riotto}, which is well-consistent with the cosmic microwave background (CMB) radiation observations.
The CMB observations have limited the amplitude of the power spectrum to the order of $\mathcal{O}(10^{-9})$ at the CMB scale~\cite{Smoot1997,Spergel2003,Komatsu2011,Akrami2020,Aghanim2020}.
 It has been found that, if the amplitude of the power spectrum of the curvature perturbations can be enhanced for about seven orders at the scales smaller than the CMB one~\cite{S. Chongchitnan,K. Kohri,C. T. Byrnes,E. Bugaev},  a sizable amount of primordial black holes can be generated  when these enhanced perturbations reenter the horizon during the radiation- or matter-dominated era~\cite{Zeldovich, Hawking, Carr, Meszaros, Carr1975, Khlopov, Ozsoy2023, Bhattacharya2023b}.

The amplitude of the power spectrum of the curvature perturbations in the standard slow-roll inflation can be expressed as $\mathcal{P_R}= \frac{H^2}{8\pi^2\epsilon c_s}$ when the mode exits the horizon during inflation, where $H$ is the Hubble parameter which is approximately constant during inflation, $\epsilon$  the slow-roll parameter and $c_s$ the sound speed of the curvature perturbations. Thus,  a natural way to  amplify the curvature perturbations is to reduce the rolling speed of the inflaton which is   proportional to $\epsilon$ or to suppress the   sound speed.   Decreasing the inflaton's rolling speed can be realized in the ultraslow-roll inflation~\cite{Yokoyama1998,Choudhury2014,Germani2017, Motohashi2017,Ezquiaga2017,H. Di2018, Ballesteros2018, Dalianis2019, Gao2018,  Tada2019, Mishra2020, Atal2020,Ragavendra2021, Bhaumik2020,   Drees2021,C.Fu2020, Xu2020, Lin2020, Dalianis2021, Yi2021,Gao2021, Yi2021b, TGao2021, Solbi2021,Gao2021b, Solbi2021b,  Zheng2021,Teimoori2021a, Cai2021, Wang2021, Fuchengjie2019,fuchengjie2020,Dalianis2020,Teimoori2021,Karam2022, Heydari2022, Heydari2022b,Bellido2017,Ezquiaga2018,Pi2022, Choudhury2023,Meng2022,Mu2022,Kawaguchi2022,Fu2022,Chen2022,Gu2023,Yi2022,Choudhury2023c,Choudhury2023e,Choudhury2023b,Cai2023, Ghoshal2023,Franciolini2023}, in which the slow-roll parameter $\eta$, which is defined to be $\eta=\frac{\dot{\epsilon}}{\epsilon H}$ with an overdot denoting a derivative with respect to the time $t$, equals approximately  $-6$.
During the transition of $\eta$ from $\eta\simeq 0$, which corresponds to the slow-roll inflation,  to $-6$, the Israel junction conditions~\cite{Israel1966,Deruelle1995}  are used to obtain the solution of the curvature perturbations in the ultraslow-roll phase. Expanding this solution in the infrared limit, one can see that  the decaying term in the solution of  the Sasaki-Mukhanov equation for the evolution of  the curvature perturbations   becomes  a growing one. 
 This growing term  gradually  dominates  if the ultraslow-roll inflation persists a sufficiently long time  and it  results in the enhancement of the curvature perturbations to meet the requirement of formation of a sizable amount of primordial black holes. It has been found that the  power spectrum of the curvature perturbations displays  a $k^4$ growth and   has a dip preceding the $k^4$ dependence~\cite{Garrilho2019,Liu2020, Byrnes2019}.  The steeper  $k^5 (\log k)^2$ growth of the power spectrum can be obtained if an $\eta=-1$ middle phase between the slow- and ultraslow-roll inflations~\cite{Garrilho2019,Cole2022} is added.

When the amplification of the curvature perturbations results from the decrease of the sound speed \cite{
G. Ballesteros,Kamenshchik,Gorji2022,Romano3,Ballesteros2022}, which equals  $1$ in the canonical scalar field inflation model,  the solution of the curvature perturbations, although does not contain  growing terms,  still has a constant component and a decay part. However,  the constant component becomes scale variant at small scales, which  makes the power spectrum  become enhanced. If the Israel junction conditions are utilized  to match the curvature perturbation and its derivative at the time when the sound speed decreases suddenly,   the power spectrum   has a $k^4$ growth~\cite{Ballesteros2022, Zhai2022}. 
However, this junction conditions is inapplicable in the case that the sound speed suddenly decreases due to the appearance of the square of the delta function~\cite{Nakashima2011}. Therefore, in this paper we will employ an improved junction conditions~\cite{Gorji2022} to restudy the evolution of the power spectrum  when the sound speed decreases suddenly and find that the growth of the  power spectrum is  $k^2$ rather than $k^4$.

Furthermore, in the Dirac-Born-Infeld-inspired nonminimal kinetic coupling inflation model~\cite{Qiu2022}, both $\epsilon$ and $c_s^2$ are closely related to the concrete form of the inflationary potential, which indicates that this inflation model may accommodate both a small sound speed of the curvature perturbations and a small rolling speed of  the inflaton at the same time. When both the inflaton's rolling speed and the sound speed of the curvature perturbations are suppressed during inflation, will the growth of the power spectrum of the curvature perturbations  be steeper than $k^4$? This is an interesting problem, which we are also going to  address in this paper.

The rest of this paper is organized as follows:
In Sec.~\ref{2},  the evolution of the power spectrum is investigated when the speed of sound decreases suddenly.
In Sec.~\ref{3}, we study the evolution of the power spectrum in the case that both the sound speed and the slow-roll parameter are suppressed  and present our conclusions in Sec.~\ref{conclusion}.
Throughout this paper, we set $c=\hbar=M_\mathrm{Pl}=1$.

\section{Growth of power spectrum from the sudden decrease of   the sound speed}
\label{2}

The evolution of the curvature perturbations $\mathcal{R}$ satisfies the Sasaki-Mukhanov equation, which in the  Fourier space takes the form
\bea\label{vk1}
v''_k+\left(c_s^2k^2-\frac{z''}{z}\right)v_k=0 \;,
 \eea
where $v_k=z\mathcal{R}_k $,  a prime indicates a derivative with respect to the conformal time $\tau$,   and  $z$ is defined as  
 \bea\label{z}
 z^2 \equiv \frac{2a^2 \epsilon}{c_s^2}
 \eea
with  $a$ being the cosmic scale factor. 
 From the definition of $z$, one can obtain \bea\label{z2}
\frac{z''}{z}=(aH)^2\left(2-\epsilon+\frac{3}{2}\eta-3s+s^2+s\epsilon-s \eta+ \frac{1}{4}\eta^2-\frac{1}{2}\eta\epsilon\right) \; .
\eea
Here
$s=\frac{\dot{c}_s}{c_s H}$.
During the (ultra)slow-roll inflation, one has $\epsilon\ll1$, and thus $aH\simeq-\frac{1}{\tau}$. Then Eq.~(\ref{vk1}) can be  rewritten as
\bea\label{vk2}
v''_k+\left(c_s^2k^2-\frac{\nu^2-1/4}{(-\tau)^2}\right)v_k=0\;,
\eea
where \bea\label{nu}
\nu \simeq \frac{3}{2}+\frac{1}{2}\eta-s\; .
\eea
If $\eta$ and $c_s$ are  constants,   Eq.~(\ref{vk2}) has a general solution
\bea\label{vk3}
v_k (\tau)=\alpha \sqrt{- \tau } H_\nu^{(1)}(-c_s k \tau )+\beta \sqrt{- \tau } H_\nu^{(2)}(-c_s k \tau )\;.
\eea
Here $H_\nu^{(1)}$ and $H_\nu^{(2)}$ are the first and second Hankel functions, respectively, and  $\alpha$ and $\beta$ are two constants.

\begin{figure}[H]
	\centering
	\includegraphics[width=0.6\linewidth]{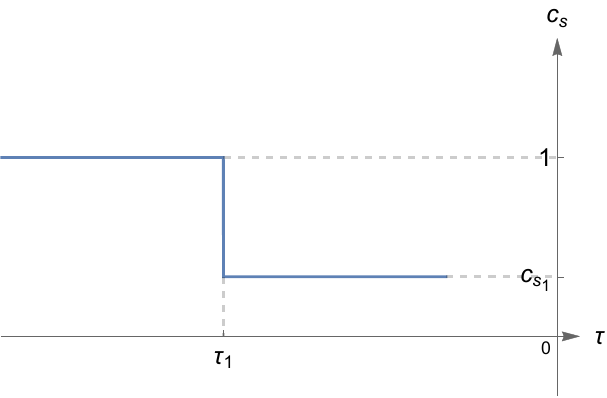}
	\caption{\label{fig1}
		The sound speed $c_s$   varies suddenly  at $\tau_1$.}
\end{figure}

\begin{figure}[H]
	\centering
\includegraphics[width=0.6\linewidth]{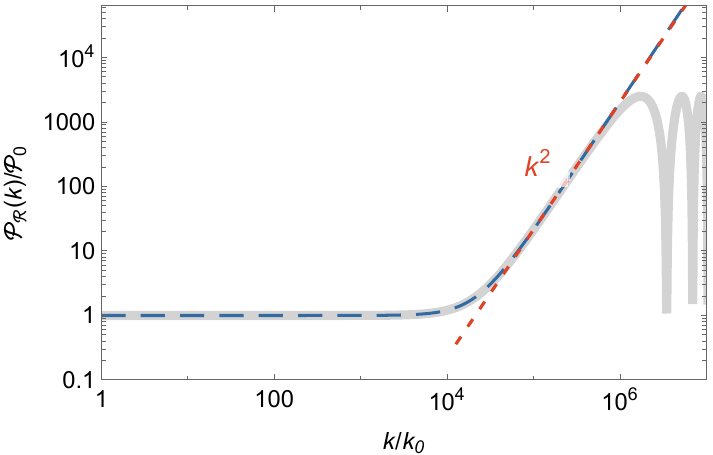}
	\caption{\label{fig2}
The evolution of the power spectrum as a function of wave number $k$. The  solid-gray and  dashed-blue lines represent the numerical and  approximate results, respectively. The dashed-red line indicates the $k^2$ growth.}
\end{figure}

Now, we discuss the scenario of a sudden decrease of the sound  speed, as shown in Fig.~\ref{fig1}.
In the first stage ($\tau<\tau_1$), which corresponds to the canonical slow-roll inflation,  the sound speed $c_s$ is equal to one
and the slow-roll parameter  $\eta$ is near zero. Thus,  one has  $\nu\simeq 3/2$ and  $\epsilon=\epsilon_0(\frac{\tau}{\tau_0})^{-\eta} \simeq \epsilon_0$. 
 Imposing that the solution of the Sasaki-Mukhanov equation  matches the plane-wave form in the  ultraviolet regime ($-k\tau\gg1$), we can derive
the evolution of the curvature perturbations \bea\label{R1}
\mathcal{R}^{(0)}_k(\tau)= i\frac{H}{2\sqrt{\epsilon_0 k^3}}e^{-i k\tau}(1+i k\tau)\;.
\eea
 Apparently, the solution of the curvature perturbations contains a constant term and a time-decaying one since $|\tau|$ decreases with the cosmic expansion during inflation,  which  results in a nearly scale-invariant power spectrum of the curvature perturbations in the superhorizon scales ($-k\tau\rightarrow 0$) with the amplitude of the power spectrum being
\bea\label{P0}
\mathcal{P}_0=\frac{H^2}{8\pi^2 \epsilon_0 }\; .
\eea
The CMB observations have limited $\mathcal{P}_0$ to be $\sim 10^{-9}$~\cite{Smoot1997}.
At the moment $\tau_1$, we assume that the sound speed  decreases suddenly from 1 to a very tiny constant value $c_{s_1}$ and the inflation enters the second stage. In this stage, 
  the general solution of the curvature perturbations has the form 
\bea\label{R-1}
\mathcal{R}^{(1)}_k (\tau)=\frac{i c_{s_1}H \left(-\tau\right)^{3/2}}{\sqrt{2\epsilon_0 }}\left[ \alpha_1 H_{3/2}^{(1)}(c_{s_1}k\tau) +\beta_1 H_{3/2}^{(2)}(c_{s_1}k\tau) \right]\;,
\eea
where $\alpha_1$ and $\beta_1$ are two constants. Usually,  the Israel junction conditions $\mathcal{R}^{(0)}_k (\tau_1)=\mathcal{R}^{(1)}_k (\tau_1)$ and $\mathcal{R}'^{(0)}_k (\tau_1)=\mathcal{R}'^{(1)}_k (\tau_1)$ are used to determine the values of $\alpha_1$ and $\beta_1$~\cite{Ballesteros2022, Zhai2022}. However, it has been found,  in Ref.~\cite{Nakashima2011},  that  there is a square term of the delta function $\delta(\tau-\tau_1)$ arising form $(c_s'/c_s)^2$ in $z''/z$ when a sudden variation of the sound velocity occurs, which makes the analysis impossible,  and thus the Israel junction conditions need to be modified or a new variable has to be introduced. In~\cite{Nakashima2011},   a new variable is defined, which satisfies the Israel junction conditions at $\tau_1$. Here, we do not use this new variable~\cite{Nakashima2011}   but consider an improved junction conditions, i.e.   $\mathcal{R}_k$ and its conjugate
momentum $A \mathcal{R}'_k$ are continuous at $\tau_1$, where $A\equiv \frac{2\epsilon}{c_{s}^2}$ ~\cite{Gorji2022}. We have checked that the introduction of the new viable and the improved junction conditions can give the same results. 
Considering the improved junction conditions
\bea \label{condition}
\mathcal{R}^{(0)}_k (\tau_1) =\mathcal{R}^{(1)}_k (\tau_1)\;,  \quad   A_0 \mathcal{R}'^{(0)}_k(\tau_1)= A_1 \mathcal{R}'^{(1)}_k (\tau_1)\;
\eea
with 
\bea
A_0= 2\epsilon_0  \;,  \quad A_1=\frac{2\epsilon_0}{c_{s_1}^2}\;,
\eea
we can obtain that
\bea
\alpha_1 &=&- \frac{\left(1-c_{s_1}\right) \sqrt{\pi}e^{-i\left(1+c_{s_1}\right)k\tau_1}}{4\sqrt{ c_{s_1}}}  ,\nonumber \\
\beta_1 &=& - \frac{\left(1+c_{s_1}\right) \sqrt{\pi}e^{-i\left(1-c_{s_1}\right)k\tau_1}}{4\sqrt{ c_{s_1}}}  \;.
\eea
Substituting $\alpha_1$ and $\beta_1$ into Eq.~(\ref{R-1}), we find the expression of the curvature perturbation in the second phase, and then we can derive the corresponding power spectrum, which is shown in Fig.~\ref{fig2}. From it, one can find that the power spectrum only has a $k^2$ growth. Thus, the result of a $k^4$ growth,   which is obtained from  the standard Israel junction conditions, should be incorrect.   

To figure out the physical  reason behind the $k^2$ growth of the power spectrum, we expand the expression of the curvature perturbations (Eq.~(\ref{R-1})) in the infrared limits: 
$- c_{s_1} k \tau\rightarrow 0$ and  $-c_{s_1} k \tau_1\rightarrow 0$, and  obtain
\bea\label{R-cs-2}
\mathcal{R}^{(1)}_k (\tau)&=&\frac{i He^{-i k\tau_1 }}{2 \sqrt{\epsilon_0 k^3}} -\frac{ H \tau_1e^{-i k\tau_1 }}{2  \sqrt{\epsilon_0 k}} -\frac{i c_{s_1}^2 H \tau_1^2  k^{1/2}  e^{-i k\tau_1} }{4 \sqrt{\epsilon_0}}+\frac{i c_{s_1}^2 H   k^{1/2}  e^{-i k\tau_1} }{4 \sqrt{\epsilon_0}}(-\tau)^2+\dots\;.
\eea
Apparently,  the wave number $k$ in Eq.~(\ref{R-cs-2}) must satisfy the condition $k\ll {k}_c\equiv -1/(c_{s_1}\tau_1)$.
It is easy to see that in the infrared limit the leading part of the curvature perturbations is   independent of $\tau$, which contains
three different $k$-dependent terms, and the subleading part decays with time since $|\tau|$ decreases during inflation.
These characters are different from that in
the case of the transition from the slow-roll inflation to the
ultraslow-roll one, where there is an appearance of the growing term. 

From Eq.~(\ref{R-cs-2}), we obtain  the  power spectrum of the curvature perturbations  
\bea\label{P-cs}
\dfrac{\mathcal{P}_{\mathcal{R}^{(1)}_k}}{\mathcal{P}_0} \simeq
1+\left(1-c_{s_1}^2\right) \tau_1^{2}k^{2}+\frac{1}{4}c_{s_1}^4\tau_1^{4}k^{4}\;
\eea
after neglecting all decaying terms. If the $k^2$ term becomes comparable to the constant one, the wave number needs to be equal to  about
\begin{eqnarray}\label{}
k_{1} \simeq - \frac{1}{\sqrt{1-c_{s_1}^2}\tau_1}\simeq c_{s_1}k_c\;.
\end{eqnarray}
The wave number at which the $k^4$ term becomes comparable with the $k^2$ one is
\begin{eqnarray}\label{}
k_{2} \simeq - \frac{2\sqrt{1-c_{s_1}^2}}{c_{s_1}^2 \tau_1}\simeq \frac{2}{c_{s_1}}k_c\;.
\end{eqnarray}
It is obvious that   $k_1\ll k_c$, but $k_2\gg k_c$ since $c_{s_1}\ll 1$. Thus, $k_2$ is beyond the  infrared condition $k\ll k_c$, which means that the power spectrum has no $k^4$ growth, and  the steepest  growth of  the power spectrum is only $k^2$.
 At the CMB scale, the first term in Eq.~(\ref{P-cs}) dominates, which leads to a scale-invariant spectrum consistent with the CMB observations. Going to the scales which are smaller than the CMB one, the second term begins to play a dominant  role. The power spectrum becomes scale dependent and has a  $k^2$ growth.
 These results are shown clearly  in Fig.~\ref{fig2}, in which the approximate result given in Eq.~(\ref{P-cs}) is very consistent with the numerical one. 
There is no dip in the power spectrum since no term cancels the constant one, which is different from the case of  the ultraslow-roll inflation.  

\section{Growth behavior of power spectrum when both the sound speed and the slow-roll parameter $\eta$ are changed suddenly}
\label{3}

We have known  that the enhancement of the power spectrum can be realized by decreasing the sound speed $c_s$ or reducing the slow-roll parameter $\epsilon$.
In the following,   we will study  the growth of the power spectrum when both $\epsilon$ and $c_s$ are suppressed. For simplicity,  we will consider that the sound speed changes suddenly from $1$ to  a constant  much less than one,  and  the slow-roll parameter $\eta$ suddenly from a constant near zero to a negative constant. A negative $\eta$ will lead to the decrease of $\epsilon$   since $\epsilon \propto \tau^{-\eta}$.  We first consider the case that the variations of  $c_s$ and $\eta$ occur simultaneously.

\subsection{Simultaneous changes of sound speed and slow-roll parameter $\eta$}
\label{sec-cs+eta}

\begin{figure}[H]
	\centering
	\includegraphics[width=0.6\linewidth]{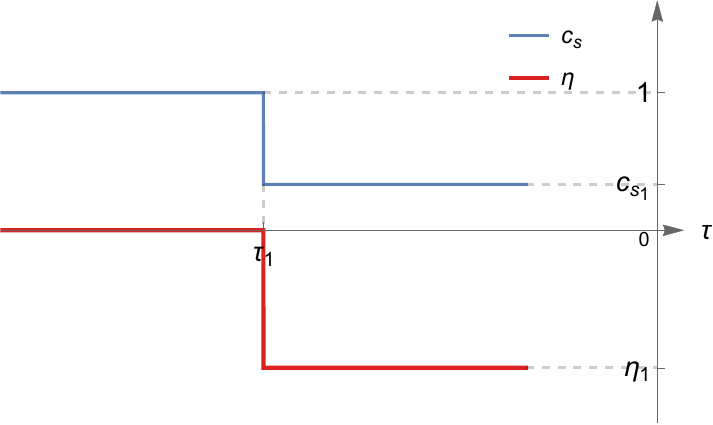}
	\caption{\label{fig3}
	 The slow-roll parameter $\eta$ (red line) and  the sound speed $c_s$ (blue line)   vary simultaneously at $\tau_1$.}
\end{figure}

The scenario considered in this subsection is shown in Fig.~\ref{fig3}. Initially, the Universe undergoes a standard slow-roll inflation, in which $c_s=1$, $\epsilon\simeq \epsilon_0\ll1$ and $\eta \sim0$. At  time $\tau_1$, the sound speed $c_s$ and the slow-roll parameter $\eta$ change suddenly from $1$ and $\sim0$ to a  small value $c_{s_1}$ and a negative constant $\eta_1$, respectively. Thus, during the second phase, the slow-roll parameter $\epsilon$ decays as $\epsilon(\tau)=\epsilon_0( \tau/\tau_1)^{-\eta_1}$ and the sound speed is a small constant $c_{s_1}$.

From Eq.~(\ref{vk3}), we obtain the general solution of the curvature perturbations
\bea\label{A-R-1}
\overline{\mathcal{R}}^{(1)}_k(\tau)=-\frac{ c_{s_1}H \tau_1^{-\frac{\eta_1}{2}}\tau^{\frac{3+\eta_1}{2}}}{\sqrt{2\epsilon_0 }}\left[ \alpha_2 H_\nu^{(1)}(c_{s_1}k\tau) +\beta_2 H_\nu^{(2)}(c_{s_1}k\tau) \right]\;
\eea
in the second phase, where $\alpha_2$ and $\beta_2$ are two constants, and $\nu=(3+\eta_1)/2$.
Matching $\mathcal{R}^{(0)}_k$ and $ \overline{\mathcal{R}}^{(1)}_k$ at $\tau= \tau_1$ by using the improved junction conditions,
$\mathcal{R}^{(0)}_k (\tau_1)=\overline{\mathcal{R}}^{(1)}_k(\tau_1)$ and $A_0 \mathcal{R}'^{(0)}_k(\tau_1)=A_1 \overline{\mathcal{R}}'^{(1)}_k (\tau_1)$,   one can achieve that
\bea \nonumber
\alpha_2 &=& \frac{\pi e^{-ik\tau_1}}{4\sqrt{2c_{s_1}^2\tau_1^3 k^3 }} \bigg[\left( \left(3+\eta_1 \right)\left(1+ik\tau_1\right)-(c_{s_1}k\tau_1)^2\right) H_{\nu}^{(2)}(c_{s_1}k\tau_1) \\  \nonumber &&  \qquad\qquad\qquad  -c_{s_1}  k \tau_1 \left(1+ik\tau_1\right) H_{\nu+1}^{(2)}(c_{s_1}k \tau_1)
\bigg]\;,  \eea
and
\bea \nonumber 
\beta_2 &=& -\frac{\pi e^{-ik\tau_1}}{4\sqrt{2c_{s_1}^2\tau_1^3 k^3 }}\bigg[\left( \left(3+\eta_1 \right)\left(1+ik\tau_1\right)-(c_{s_1}k\tau_1)^2\right) H_{\nu}^{(1)}(c_{s_1}k\tau_1)  \\ && \nonumber 
\qquad\qquad\qquad -c_{s_1}  k \tau_1 \left(1+ik\tau_1\right) H_{\nu+1}^{(1)}(c_{s_1}k \tau_1)
\bigg] \;.
\eea
Substituting $\alpha_2$ and $\beta_2$ into Eq.~(\ref{A-R-1}) gives the expression of the curvature perturbations    during the second phase.
Expanding this expression in the infrared limit ($-c_{s_1} k \tau \rightarrow 0$
 and $-c_{s_1} k \tau_1 \rightarrow 0$),
we arrive at
{\small
 \begin{eqnarray}\label{A-R-1-IR}
  	\overline{\mathcal{R}}^{(1)}_k(\tau)&=&
  	\frac{i  H e^{-i k\tau_1 }}{2 \sqrt{\epsilon_0 k^3}}-
  	\frac{ H \tau_1e^{-i k \tau_1}}{2 \sqrt{ \epsilon_0 k}}
-
  	\left( \frac{3 i c_{s_1}^{2}  H  \tau_1^2e^{-i k\tau_1}}{4 \sqrt{\epsilon_0}(3+\eta_1)}
  		+ \frac{i c_{s_1}^{2}  H \eta_1  \left(-\tau_1\right)^{-1-\eta_1}e^{-i k\tau_1}}{2 \sqrt{\epsilon_0}(1+\eta_1)(3+\eta_1) }(-\tau)^{3+\eta_1}
  	\right)k^{1/2}\nonumber\\
  	&+&\left(\frac{c_{s_1}^2  H \tau_1^3 e^{-i k \tau_1}}{4 \sqrt{\epsilon_0}(3+\eta_1)}-
  	\frac{ c_{s_1}^{2}  H e^{-i k \tau_1}}{2 \sqrt{\epsilon_0}(1+\eta_1)(3+\eta_1)\left(-\tau_1\right)^{\eta_1} }(-\tau)^{3+\eta_1}
  	\right)k^{3/2}
  	\nonumber\\
   &+&\dots\;.
  \end{eqnarray}	}
It is easy to see that the solution contains a time-independent part and a time-dependent one,
which will decay with time when $\eta_1>-3$, and grow  when $\eta_1<-3$.

Now, we assume that the second phase lasts for $N_2$ number of e-folds, which means that $\tau_2=\tau_1 e^{-N_2}$ if this phase ends at $\tau=\tau_2$.
The power spectrum of the curvature perturbations can be obtained from Eq.~(\ref{A-R-1-IR}), which takes the form
{\small
	\begin{eqnarray}\label{A-P-IR}
		\dfrac{\mathcal{P}_{\overline{\mathcal{R}}^{(2)}_k}}{\mathcal{P}_0} &\simeq&
		1+\left(1+\frac{2c_{s_1}^{2} \eta_1}{(1+\eta_1)(3+\eta_1)} e^{-\left(3+\eta_1 \right)N_2}\right)\tau_1^2k^2
		\nonumber\\	
				&-&\left(\begin{split}&
			\frac{1}{3+\eta_1}+\frac{2 }{(1+\eta_1)(3+\eta_1)} e^{-\left(3+\eta_1 \right)N_2}
			-
			\frac{c_{s_1}^{2}\eta_1^2 }{(1+\eta_1)^2(3+\eta_1)^2} e^{-2\left(3+\eta_1 \right)N_2}
		\end{split}
			\right)c_{s_1}^{2} \tau_1^4k^4
		\nonumber\\
		&+&
		\left(\begin{split}
			&\frac{1}{4(3+\eta_1)^2}+\frac{1 }{(1+\eta_1)(3+\eta_1)^2} e^{-\left(3+\eta_1 \right)N_2}
+
			\frac{1 }{(1+\eta_1)^2(3+\eta_1)^2} e^{-2\left(3+\eta_1 \right)N_2}
		\end{split}\right) c_{s_1}^4 \tau_1^6k^6 	\;.
		\nonumber\\
\end{eqnarray}}
Here  we have kept the $k^6$ term in Eq.~(\ref{A-P-IR}) since the $k^6$ growth may occur under certain conditions. We need to  compare the $k^n$ term with the $k^{n-2}$ one to determine whether a $k^n$ growth will appear.
We set $k_1$, $k_2$, and $k_3$ to denote the wave number  at which the scale-invariant term becomes comparable with the $k^2$ term, the $k^2$ term becomes comparable with the $k^4$ term, and the $k^4$ term becomes comparable with the $k^6$ term, respectively.

\begin{figure}[H]
	\centering
\includegraphics[width=0.6\linewidth]{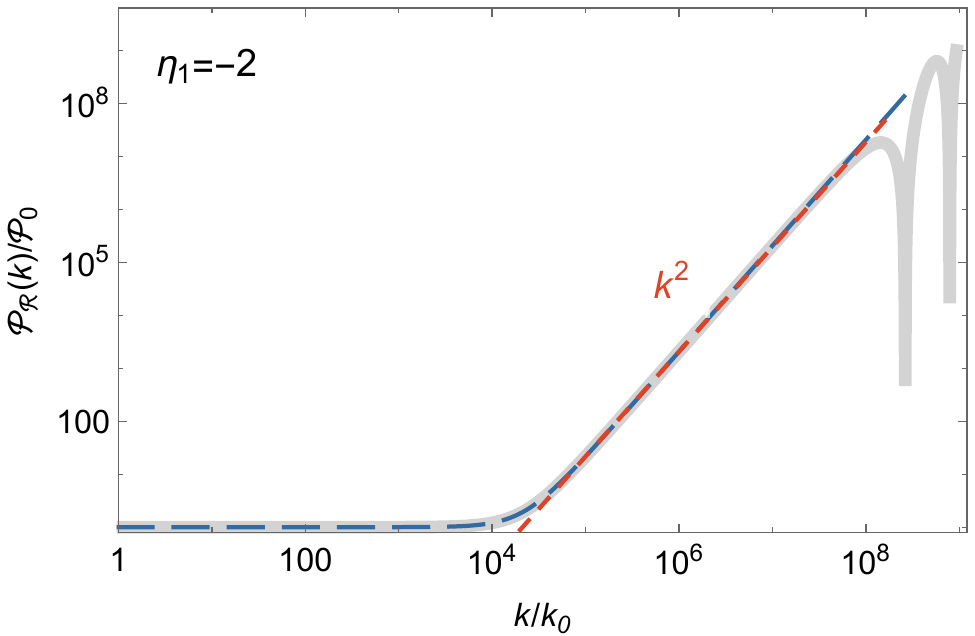}
	\caption{\label{fig4}
The   power spectrum in the $\eta_1=-2$ case. The gray-solid and  blue-dashed lines represent the numerical  and approximate results, respectively. The dashed-red line indicates the $k^2$ growth.}
\end{figure}

Let us first study the $\eta_1>-3$ case, which means that there are no growing terms in the solution of the curvature perturbations and all time-dependent terms in Eq.~(\ref{A-R-1-IR}) decay with the cosmic expansion. Thus,  all terms containing $N_2$ in Eq.~(\ref{A-P-IR}) can be neglected.
Then, we find that  
\begin{eqnarray}\label{k}
k_{1} \simeq -\frac{1}{\tau_1}=c_{s_1}k_c \;, \end{eqnarray}
and 
\begin{eqnarray}
k_{2} \simeq -\frac{\sqrt{3+\eta_1}}{c_{s_1}\tau_1}=\sqrt{3+\eta_1}k_c \;.
\end{eqnarray}
The condition $k_{2}\ll k_c$ for a $k^4$ growth requires that $ \sqrt{3+\eta_1}\ll1$. This is hard to satisfy since $\eta_1$ must be fine-tuned to be very close to $-3$.
Thus,   usually the highest growth of the power spectrum can only reach $k^2$.
Furthermore, since the power spectrum only has the $k^2$ growth, the dip phenomenon does not appear.
These characters can be found clearly in Fig.~\ref{fig4}, where the  power spectrums from numerical and approximate analyses are plotted in the $\eta_1=-2$ case.

\begin{table}[htbp]
	\caption{ The predominant terms of $k^n$ in three different situations. }\label{tab1}
	\renewcommand\arraystretch{1.4}
	\setlength{\tabcolsep}{7mm}
	\begin{tabular}{c|c|c|c}
		\hline
		\hline
		& $c_{s_1}^2e^{-(3+\eta_1)N_2}\ll 1$           & $c_{s_1}^2e^{-(3+\eta_1)N_2}= 1$    & $c_{s_1}^2e^{-(3+\eta_1)N_2}\gg 1$\\
		\hline
	$\tau_1^2 k^2$ 	&  $1$  & $1+\frac{2 \eta_1}{(1+\eta_1)(3+\eta_1)} $ & $\frac{2c_{s_1}^{2} \eta_1 e^{-(3+\eta_1)N_2}}{(1+\eta_1)(3+\eta_1)}$ \\
		\hline
		$c_{s_1}^{2}\tau_1^4 k^4$	& $-\frac{2e^{-\left(3+\eta_1 \right)N_2} }{(1+\eta_1)(3+\eta_1)} $
		& $-\frac{(\eta_1^2+8\eta_1+6) e^{-\left(3+\eta_1 \right)N_2}}{(1+\eta_1)^2(3+\eta_1)^2} $ & $
			\frac{c_{s_1}^{2}\eta_1^2 e^{-2\left(3+\eta_1 \right)N_2} }{(1+\eta_1)^2(3+\eta_1)^2} $   \\
		\hline
	$c_{s_1}^4 \tau_1^6 k^6$	&  $
			\frac{e^{-2\left(3+\eta_1 \right)N_2} }{(1+\eta_1)^2(3+\eta_1)^2}
	$
		&
			$\frac{e^{-2\left(3+\eta_1 \right)N_2} }{(1+\eta_1)^2(3+\eta_1)^2}
			$ & $\frac{e^{-2\left(3+\eta_1 \right)N_2} }{(1+\eta_1)^2(3+\eta_1)^2} $ \\
		\hline
		\hline
	\end{tabular}
\end{table}

When $\eta_1<-3$,  since the coefficient of the $k^n$ term is very tedious,  we only consider  three special cases ($c_{s_1}^2e^{-(3+\eta_1)N_2}\ll 1$, $c_{s_1}^2e^{-(3+\eta_1)N_2}= 1$ and $c_{s_1}^2e^{-(3+\eta_1)N_2}\gg 1$),  as shown in Table~\ref{tab1} in which  the predominant terms  of $k^n$ are given, to investigate analytically the growth of the power spectrum.
In the case of $c_{s_1}^2e^{-(3+\eta_1)N_2}\ll 1$, we find that 
\begin{eqnarray}\label{}
	k_{1}  &\simeq&-\frac{1}{  \tau_1}=c_{s_1}k_c \;,\nonumber\\
    k_{2}  &\simeq&-\frac{\sqrt{(1+\eta_1)(3+\eta_1)}}{ \sqrt{2} c_{s_1} \tau_1} e^{\frac{(3+\eta_1)N_2}{2}}= \frac{\sqrt{(1+\eta_1)(3+\eta_1)}}{ \sqrt{2} } e^{\frac{(3+\eta_1)N_2}{2}} k_c\;,\nonumber\\
    k_{3}  &\simeq&-\frac{\sqrt{2(1+\eta_1)(3+\eta_1)}}{  c_{s_1} \tau_1} e^{\frac{(3+\eta_1)N_2}{2}}=\sqrt{2(1+\eta_1)(3+\eta_1)} e^{\frac{(3+\eta_1)N_2}{2}}k_c \;.
\end{eqnarray}
Apparently,  the condition  $ k_3 < k_c$ is easy to be satisfied since $\eta_1<-3$. Thus the power spectrum can have a $k^6$ growth.  Since the coefficient of the $k_4$ term is negative, the power spectrum has a dip preceding the  $k^6$ growth.

When $c_{s_1}^2e^{-(3+\eta_1)N_2}= 1$, we have
\begin{eqnarray}\label{}
	k_{1}  &\simeq& \sqrt{\frac{\eta_1^2+4\eta_1+3}{(\eta_1^2+6\eta_1+3)}} c_{s_1}k_c\;,\nonumber\\
    k_{2}  &\simeq& \sqrt{\frac{(\eta_1^2+4\eta_1+3)(\eta_1^2+6\eta_1+3)}{(\eta_1^2+8\eta_1+6)}} c_{s_1}k_c\;,\nonumber\\
    k_{3}  &\simeq& \sqrt{\eta_1^2+8\eta_1+6} ~c_{s_1}k_c\;.
\end{eqnarray}
We find that $k_1\simeq k_2\simeq k_3<k_c$,  which means that the power spectrum will go directly to the $k^6$ growth after the scale-invariant spectrum.

\begin{figure}[H]
	\centering
\includegraphics[width=0.3\linewidth]{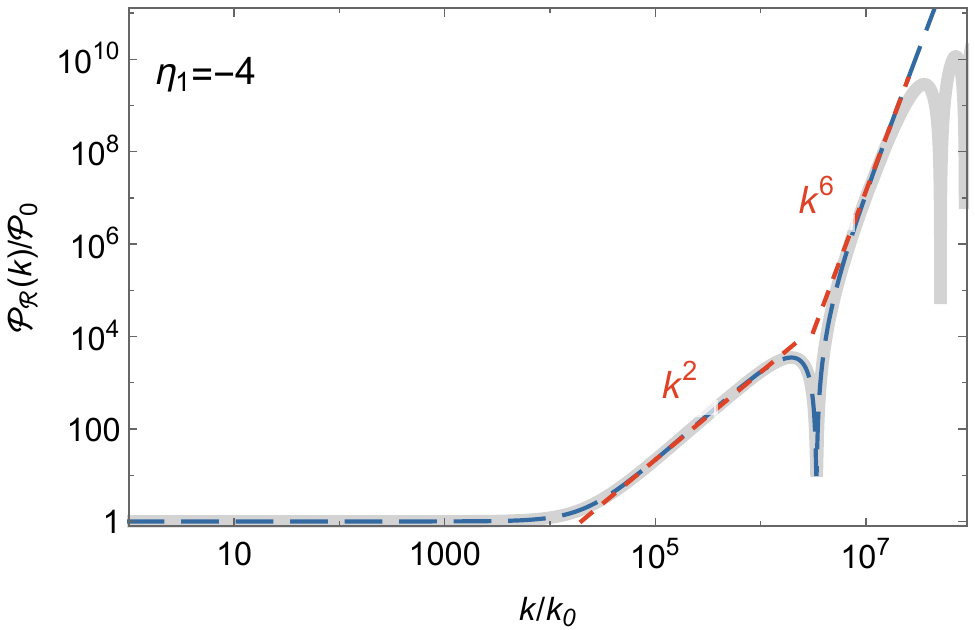}
  \includegraphics[width=0.3\linewidth]{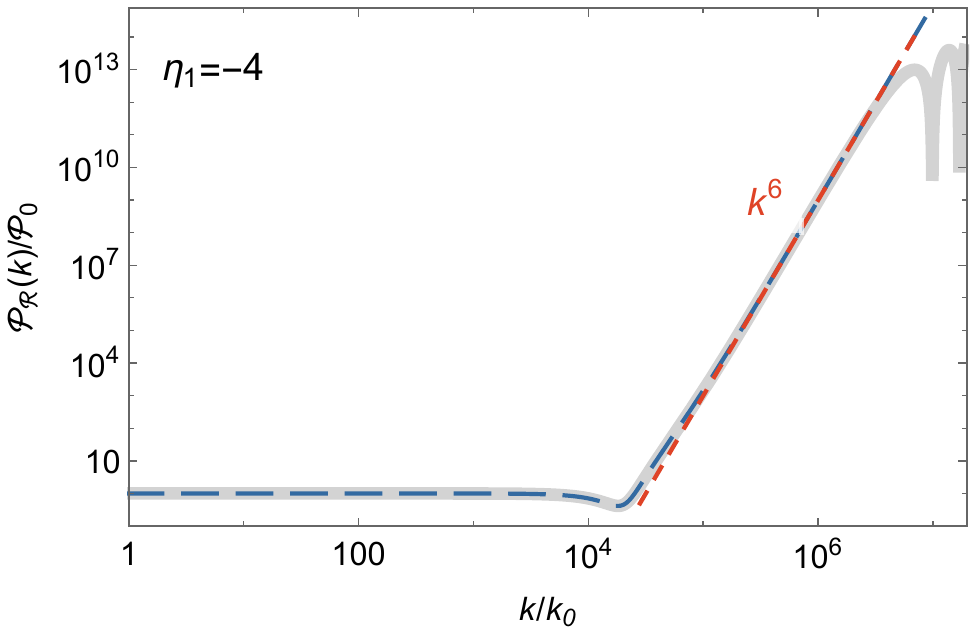}
   \includegraphics[width=0.3\linewidth]{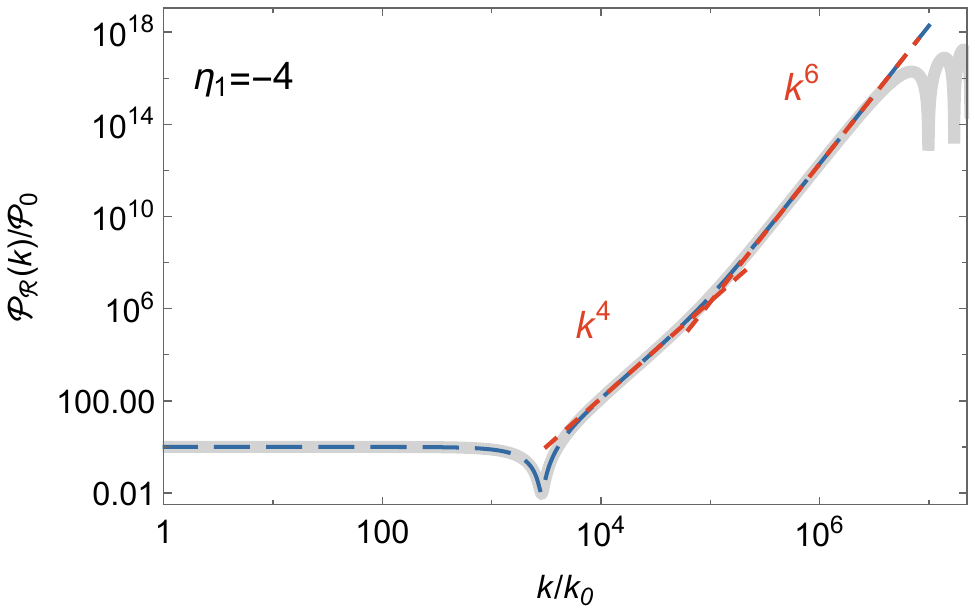}
 \includegraphics[width=0.3\linewidth]{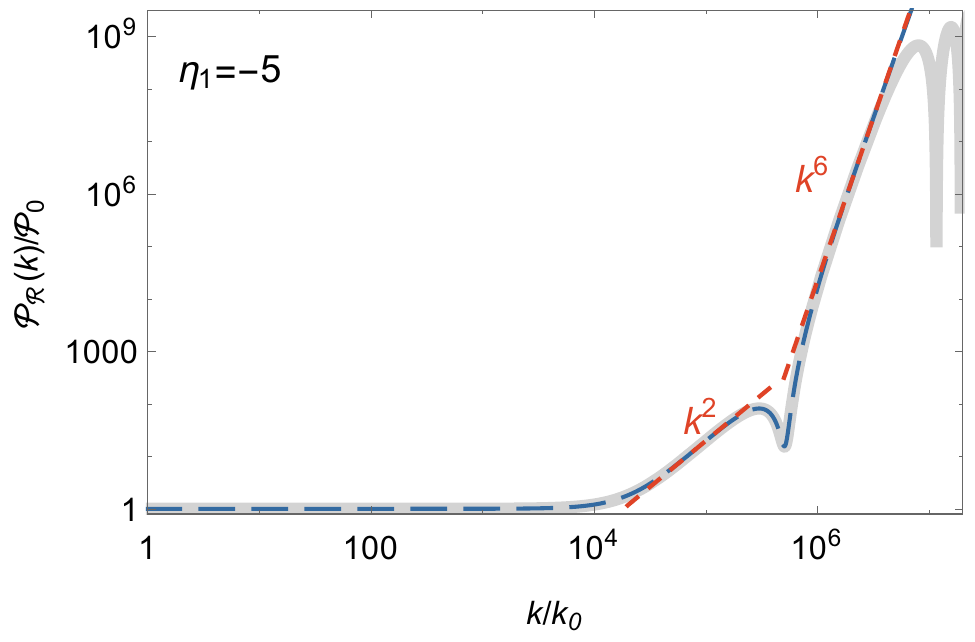}
  \includegraphics[width=0.3\linewidth]{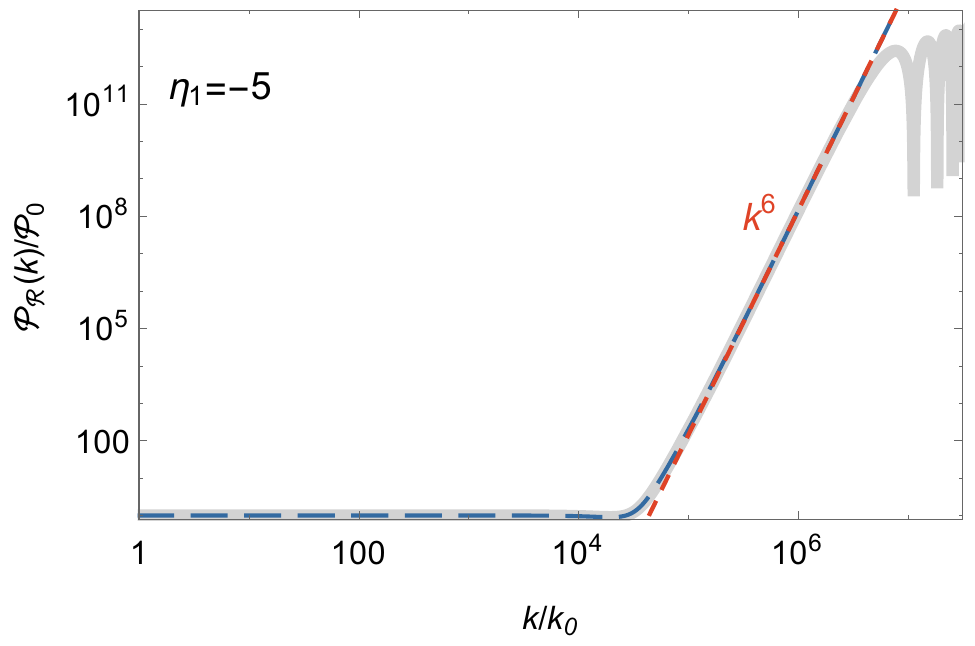}
   \includegraphics[width=0.3\linewidth]{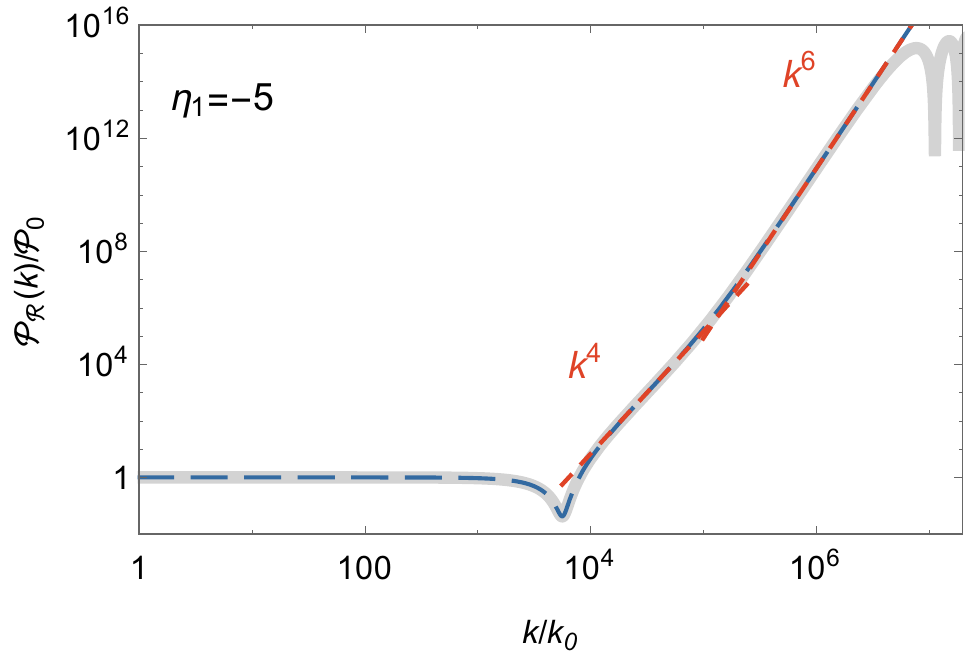}
 \includegraphics[width=0.3\linewidth]{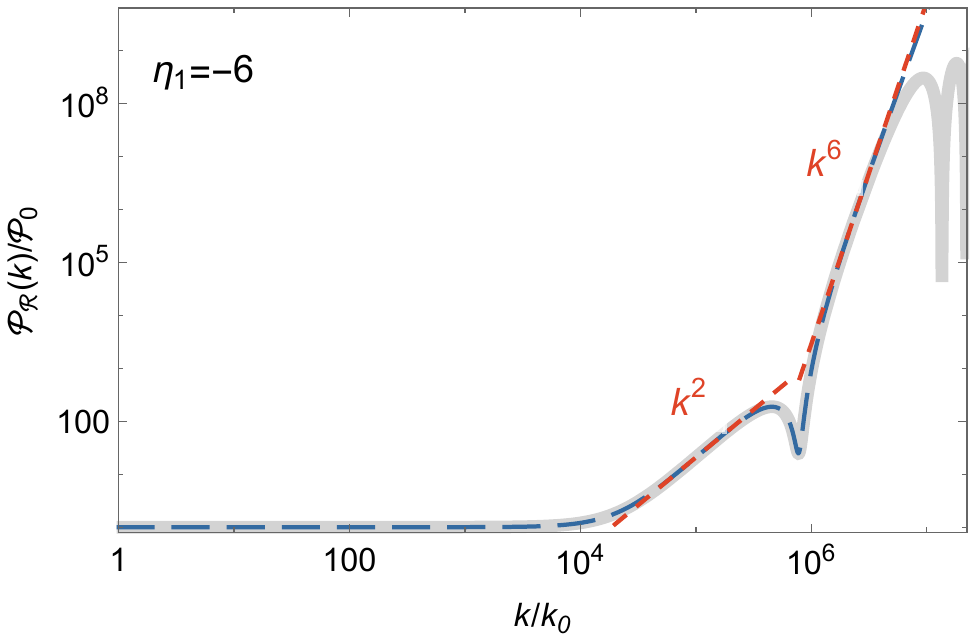}
  \includegraphics[width=0.3\linewidth]{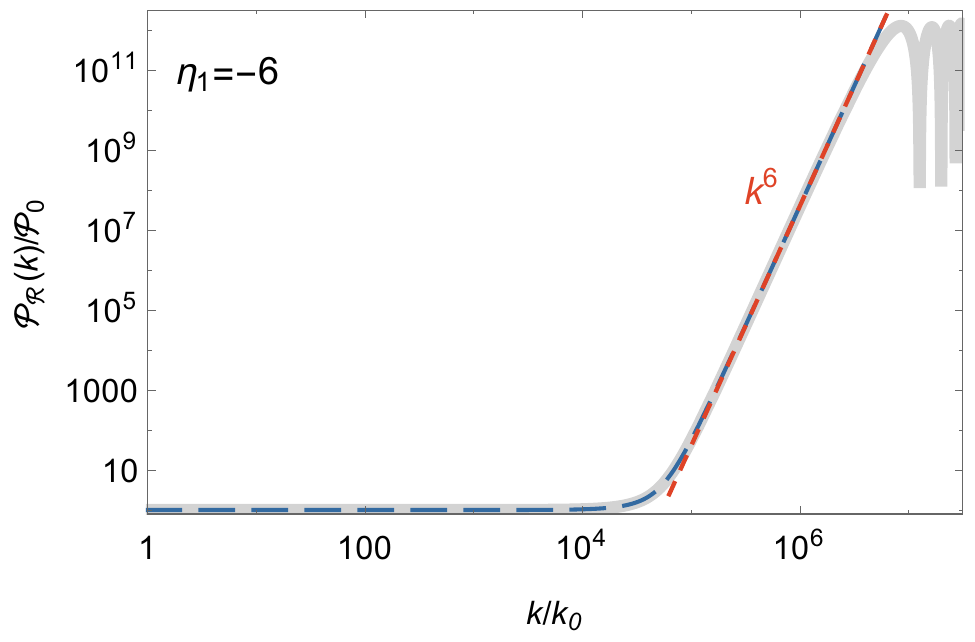}
   \includegraphics[width=0.3\linewidth]{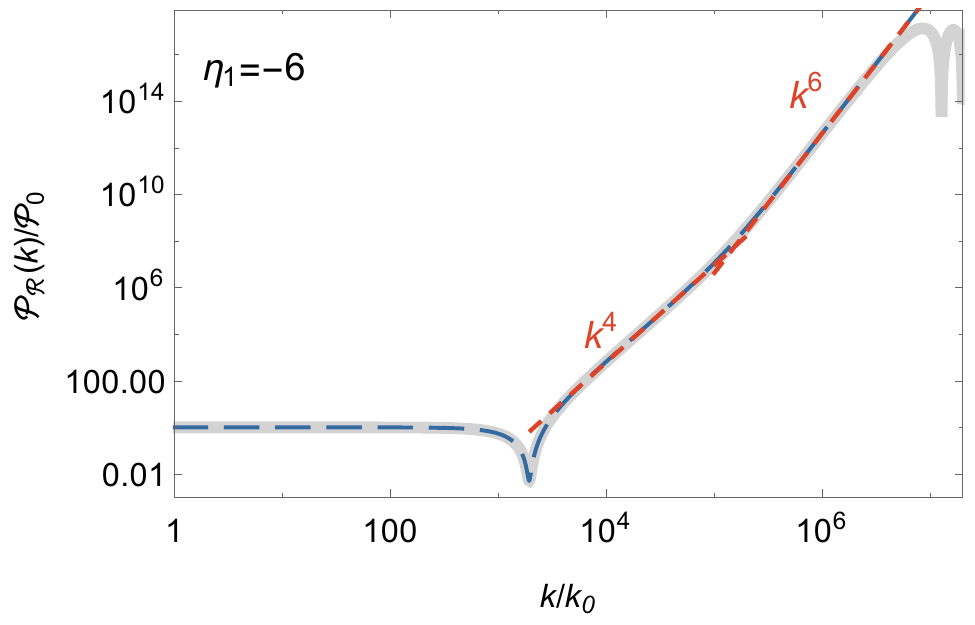}
	\caption{\label{fig5}
The  power spectrums in the $\eta_1=-4$, $-5$, and $-6$ cases. The gray-solid and  blue-dashed lines represent the numerical and approximate   results, respectively.
The first, middle and last  columns correspond to  the $c_{s_1}^2e^{-(3+\eta_1)N_2}\ll 1$,   $c_{s_1}^2e^{-(3+\eta_1)N_2}= 1$, and $c_{s_1}^2e^{-(3+\eta_1)N_2}\gg 1$ cases, respectively.}
\end{figure}

In the case of $c_{s_1}^2e^{-(3+\eta_1)N_2}\gg 1$, one can obtain 
\begin{eqnarray}\label{}
	k_{1}  &\simeq& \frac{\sqrt{(1+\eta_1)(3+\eta_1)}}{ \sqrt{-2\eta_1} } e^{\frac{1}{2}(3+\eta_1)N_2} k_c\;,\nonumber\\
    k_{2}  &\simeq& \frac{\sqrt{(1+\eta_1)(3+\eta_1)}}{ \sqrt{-\eta_1} } e^{\frac{1}{2}(3+\eta_1)N_2} k_c \;,\nonumber\\
    k_{3}  &\simeq&  |\eta_1| c_{s_1}k_c\;.
\end{eqnarray}
Since  $ k_2 \ll k_c$,  the power spectrum will
grow  with $k^4$  when the scales become smaller than the CMB one. It eventually goes into $k^6$ growth due to $k_3<k_c$. A negative $\eta_1$ leads to a negative coefficient of the $k^2$ term, which indicates that the power spectrum has a dip preceding the $k^4$ growth.  These characters can be seen in Fig.~\ref{fig5}, where the numerical   (gray -solid lines) and approximate (blue-dashed  lines) results  of the power spectrum with $\eta_1=-4,-5,-6$ are plotted.  One can see that the approximate results are consistent with the numerical ones.
 
\begin{figure}[H]
	\centering
\includegraphics[width=0.6\linewidth]{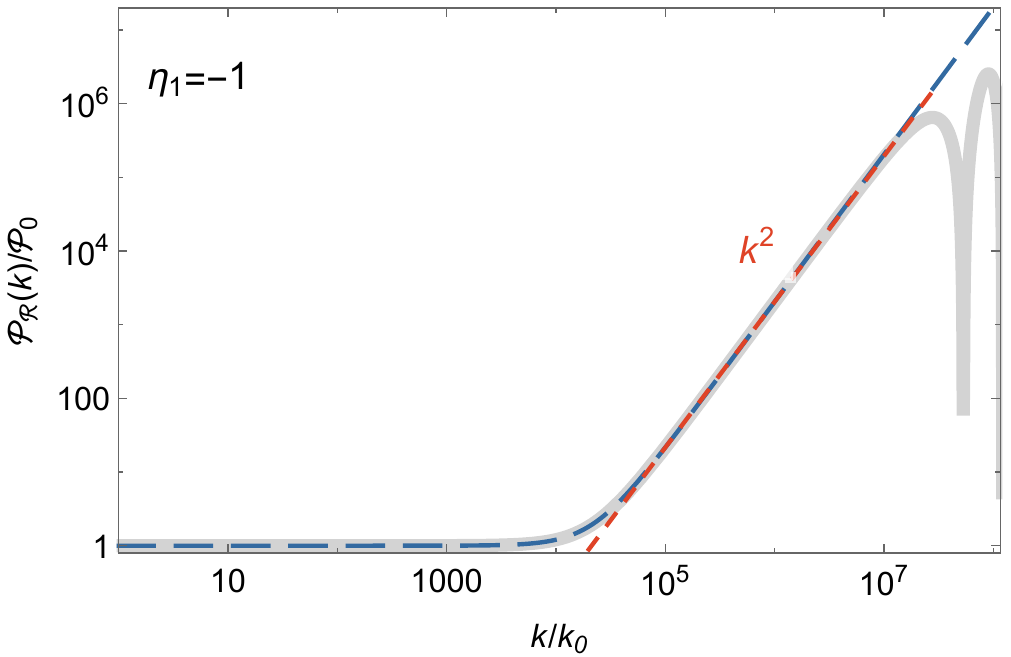}
	\caption{\label{fig6}
The   power spectrum in the $\eta_1=-1$ case .The gray-solid and  blue-dashed lines represent the numerical  and approximate results, respectively. The dashed-red line indicates the $k^2$ growth.}
\end{figure}

 Moreover, Eq.~(\ref{A-R-1-IR}) is clearly inapplicable for the cases of $\eta_1=-1$ and $\eta_1=-3$ due to the appearance of singularity.
These two cases need to be treated separately.
\begin{itemize}
	\item $\eta_1=-1$
\end{itemize}
We find that in the infrared limit, the solution of the curvature perturbations [Eq.~(\ref{A-R-1})] has the form
\begin{eqnarray}\label{}
	\overline{\mathcal{R}}^{(1)}_k(\tau) =
	\frac{i He^{-i k\tau_1 }}{2 \sqrt{\epsilon_0 k^3}}-
	\frac{ H \tau_1e^{-i k \tau_1 }}{2 \sqrt{ \epsilon_0 k}}-
	\dfrac{ic_{s_1}^2 H  k^{1/2} e^{-i k\tau_1 } }{8  \sqrt{\epsilon_0 }}\left[3\tau_1^2-(-\tau)^2\right]
	+\dots \;.
\end{eqnarray}
The solution of the curvature perturbations consists of the constant part and the decaying  one, which is similar to the solution in   the case of $\eta_1>-3$.
The  power spectrum of the  curvature perturbations has the form
\bea\label{B-P-1-ir-1}
\dfrac{\mathcal{P}_{\overline{\mathcal{R}}^{(1)}_k}}{\mathcal{P}_0} \simeq
1 + \tau_1^{2}k^{2}
+\frac{9}{16}c_{s_1}^{4}\tau_1^{4}k^{4}\;.
\eea
We can see  that
the $k^2$ and $k^4$ terms  become comparable at
\begin{eqnarray}\label{}
k_{2}\approx -\frac{4}{3c_{s_1}^2\tau_1}= \frac{4}{3c_{s_1}}k_c \;.
\end{eqnarray}
Since $k_2>k_c$, the power spectrum has no $k^4$ growth.
The corresponding numerical and approximate results of the power
spectrum are shown in Fig.~\ref{fig6}. Apparently at the small scales the power spectrum grows with
a $k^2$ slope.

\begin{figure}[H]
	\centering
\includegraphics[width=0.6\linewidth]{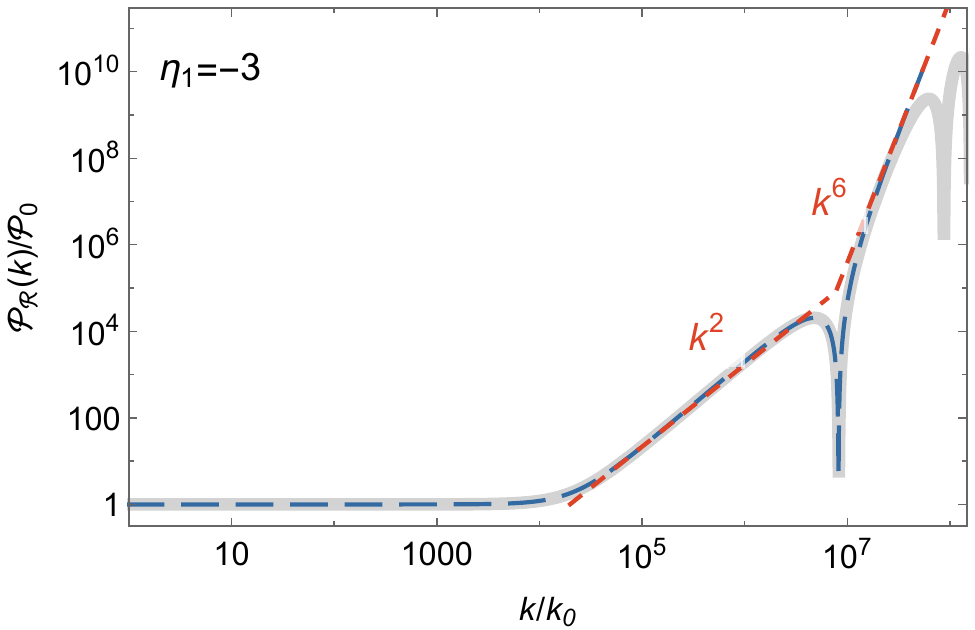}
	\caption{\label{fig7}
The   power spectrum in the $\eta_1=-3$ case. The gray-solid and  blue-dashed lines represent the numerical  and approximate results, respectively. }
\end{figure}

\begin{itemize}
	\item $\eta_1=-3$
\end{itemize}
In the infrared limit, Eq.~(\ref{A-R-1}) can be simplified to be
\begin{eqnarray}\label{}
	\overline{\mathcal{R}}^{(1)}_k(\tau) &=&
	\frac{i He^{-i k\tau_1 }}{2 \sqrt{\epsilon_0 k^3}}-
	\frac{ H \tau_1e^{-i k \tau_1 }}{2 \sqrt{ \epsilon_0 k}}+
	\dfrac{ic_{s_1}^2 H \tau_1^2 k^{1/2} e^{-i k\tau_1 } }{8  \sqrt{\epsilon_0 }}\left[1+6\log \left( \dfrac{\tau}{\tau_1}\right) \right]\nonumber\\
	&-&\dfrac{ic_{s_1}^2 H \tau_1^3 k^{3/2} e^{-i k\tau_1 } }{8  \sqrt{\epsilon_0 }}\left[1+2\log \left( \dfrac{\tau}{\tau_1}\right) \right]
	+\dots \;.
\end{eqnarray}
The solution for the curvature perturbation consists of the constant term and the logarithmically growing one. This characteristic is different from the power-law growth in the $\eta_1<-3$ case.
Thus, we obtain that the power spectrum has the  expression
\bea\label{Eq29}
\dfrac{\mathcal{P}_{\overline{\mathcal{R}}^{(1)}_k}}{\mathcal{P}_0} &\simeq&
1+\frac{1}{2}\left(2-c_{s_1}^{2}\left(6N_2-1\right)\right) \tau_1^{2}k^{2}
+\frac{1}{16}c_{s_1}^{2}\left(8+c_{s_1}^{2}\left(6N_2-1\right)^2 -16N_2\right) \tau_1^{4}k^{4}
\nonumber\\&+&
\frac{1}{16}c_{s_1}^{4}\left(2N_2-1\right)^2 \tau_1^{6}k^{6}
\;.
\eea
Since the maximum value of  $N_2$ is about  $30-40$  and  $c_{s_1}\sim \mathcal{O} (10^{-4})$,  $N_2 c_{s_1}^{2}$ is significantly less than 1.
Therefore, the expression of the power spectrum given in Eq.~(\ref{Eq29}) can be simplified to be  
\bea\label{}
\dfrac{\mathcal{P}_{\overline{\mathcal{R}}^{(1)}_k}}{\mathcal{P}_0} \simeq
1 + \tau_1^{2}k^{2}
+\frac{1}{2}(1-2N_2) c_{s_1}^{2} \tau_1^{4}k^{4}
+
\frac{1}{16}(2N_2-1)^2 c_{s_1}^{4} \tau_1^{6}k^{6}
\;.
\eea
From the above expression, we obtain 
\begin{eqnarray}\label{k_62}
	k_{1}  &\simeq& c_{s_1}k_c \;,\nonumber\\
    k_{2}  &\simeq& \frac{1}{ \sqrt{N_2-1/2 }} k_c \;,\nonumber\\
    k_{3}  &\simeq& \frac{2}{\sqrt{N_2-1/2}} k_c \;.
\end{eqnarray}
Obviously, $k_1 \ll k_c$  and $k_2 \approx k_3$, which are less than  $k_c$ if $N_2 >9/2$. Thus,  the power spectrum   will have an  era  with a  $k^2$ growth and eventually with a $k^6$  one at scales smaller than the CMB one when $N_2 >9/2$.  Since the coefficient of the $k^4$ term is negative, there is a dip preceding the $k^6$ growth. These characters can be seen clearly in Fig.~\ref{fig7}.

\subsection{Changes of sound speed previous to slow-roll parameter $\eta$}
\label{sec-cs-eta}
\begin{figure}[H]
	\centering
	\includegraphics[width=0.6\linewidth]{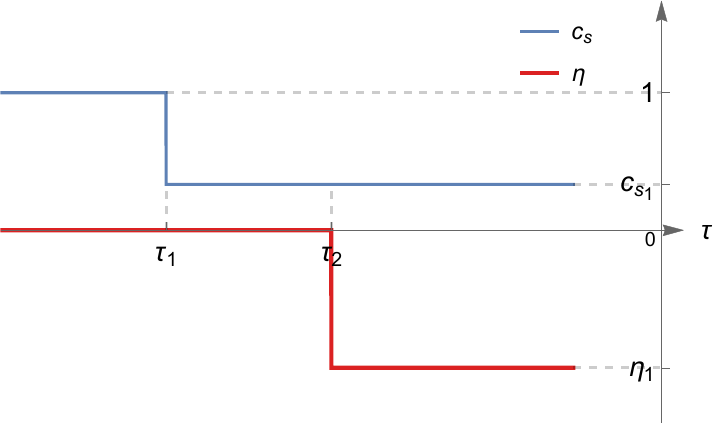}
	\caption{\label{fig8}
  The sound speed $c_s$ (blue line)   varies  at $\tau_1$, and the slow-roll parameter $\eta$ (red line) changes at $\tau_2$   $(|\tau_2|<|\tau_1|$).
	}
\end{figure}

The simultaneous change of $c_s$ and $\eta$ is a harsh requirement. In the following, we abandon it, and first  consider that the change of the sound speed is followed by  that of $\eta$. We assume that the sound speed changes suddenly  from $1$ to a  small constant $c_{s_1}$  at time $\tau_1$, and  $\eta$ varies from $\sim 0$ to a negative constant $\eta_1$ at $\tau_2$ ($|\tau_2|<|\tau_1|$), as is shown in Fig.~\ref{fig8}. The $|\tau|>|\tau_1|$ era, which represents  the first phase, is the standard slow-roll inflation. In the second phase ($|\tau_2|<|\tau|<|\tau_1|$), the sound speed of the curvature perturbations is a small value $c_{s_1}$. When $|\tau|<|\tau_2|$, which corresponds to the third phase, the  slow-roll parameter $\epsilon$ decreases with the power-law: $\epsilon(\tau)=\epsilon_0(\tau/\tau_2)^{-\eta_1}$ and the sound speed keeps the small value $c_{s_1}$.

When $|\tau|>|\tau_1|$, the solution of $\mathcal{R}_k^{(0)}$ is given in Eq.~(\ref{R1}).
In the second phase,  the solution of the curvature perturbations has the form
\bea\label{RB2-R}
\widetilde{\mathcal{R}}^{(1)}_k (\tau)=\frac{i c_{s_1}H \left(-\tau\right)^{3/2}}{\sqrt{2\epsilon_0 }}\left[ \alpha_3 H_{3/2}^{(1)}(c_{s_1}k\tau) +\beta_3 H_{3/2}^{(2)}(c_{s_1}k\tau) \right] \;,
\eea
where $ \alpha_3$ and $\beta_3$ are two constants. Using the matching condition: $\mathcal{R}^{(0)}_k(\tau_1)=\widetilde{\mathcal{R}}^{(1)}_k(\tau_1)$ and $A_0\mathcal{R}'^{(0)}_k(\tau_1)=A_1 \widetilde{\mathcal{R}}'^{(1)}_k(\tau_1)$, one can obtain that
\bea
\alpha_3  &=&- \frac{\left(1-c_{s_1}\right) \sqrt{\pi}e^{-i\left(1+c_{s_1}\right)k\tau_1}}{4\sqrt{ c_{s_1}}}  ,\nonumber \\
\beta_3 &=& - \frac{\left(1+c_{s_1}\right) \sqrt{\pi}e^{-i\left(1-c_{s_1}\right)k\tau_1}}{4\sqrt{ c_{s_1}}}  \;.
\eea
If $|\tau|<|\tau_2|$,
the solution of the curvature perturbations becomes
\bea\label{RB3-R}
\widetilde{\mathcal{R}}^{(2)}_k(\tau)=-\frac{ c_{s_1}H \tau_2^{-\frac{\eta_1}{2}}\tau^{\frac{3+\eta_1}{2}}}{\sqrt{2\epsilon_0 }}\left[ \alpha_4 H_\nu^{(1)}(c_{s_1}k\tau) +\beta_4 H_\nu^{(2)}(c_{s_1}k\tau) \right],
\eea
where $\alpha_4$ and $\beta_4$ are two constants and $\nu=(3+\eta_1)/2$.
Using the matching condition: $\widetilde{\mathcal{R}}^{(1)}_k(\tau_2)=\widetilde{\mathcal{R}}^{(2)}_k(\tau_2)$ and $\widetilde{\mathcal{R}}'^{(1)}_k(\tau_2)= \widetilde{\mathcal{R}}'^{(2)}_k(\tau_2)$,  one finds
\begin{eqnarray}
	\alpha_4 &=& \frac{i \pi^{3/2}c_{s_1}^{1/2}\tau_2 k}{16 }e^{-i(1+c_{s_1})k\tau_1}
	\left[~\xi H_{\frac{3+\eta_1}{2}}^{(2)}(c_{s_1}k\tau_2)
	-
	\lambda H_{\frac{1+\eta_1}{2}}^{(2)}(c_{s_1}k\tau_2)  \right],
	\\
	\beta_4 &=&-\frac{i \pi^{3/2}c_{s_1}^{1/2}\tau_2 k}{16 }e^{-i(1+c_{s_1})k\tau_1}
	\left[~\xi H_{\frac{3+\eta_1}{2}}^{(1)}(c_{s_1}k\tau_2)
	-
	\lambda H_{\frac{1+\eta_1}{2}}^{(1)}(c_{s_1}k\tau_2)  \right]
\;,
\end{eqnarray}
where
\begin{align}
	\xi&=(1-c_{s_1})H_{\frac{1}{2}}^{(1)}(c_{s_1}k\tau_2)+(1+c_{s_1})e^{2ic_{s_1}k\tau_1}H_{\frac{1}{2}}^{(2)}(c_{s_1}k\tau_2)\;,\nonumber\\
	\lambda&=(1-c_{s_1})H_{\frac{3}{2}}^{(1)}(c_{s_1}k\tau_2)+(1+c_{s_1})e^{2ic_{s_1}k\tau_1}H_{\frac{3}{2}}^{(2)}(c_{s_1}k\tau_2)\;.
\end{align}
In the infrared region ($-c_{s_1} k \tau \rightarrow 0$, $-c_{s_1} k \tau_1 \rightarrow 0$ and $-c_{s_1} k \tau_2 \rightarrow 0$ ), we obtain that
   \begin{eqnarray}\label{B-R-IR}
  	\widetilde{\mathcal{R}}^{(2)}_k(\tau)&=&
  	\frac{i  H e^{-i k\tau_1 }}{2 \sqrt{\epsilon_0 k^3}}-
  	\frac{ H \tau_1e^{-i k \tau_1}}{2 \sqrt{ \epsilon_0 k}}\nonumber\\
  &-&
  	\left( \frac{i  c_{s_1}^{2}  H \left((3+\eta_1)\tau_1^2-\eta_1\tau_2^2\right)e^{-i k\tau_1}}{4 \sqrt{\epsilon_0}(3+\eta_1)}
  		+ \frac{ic_{s_1}^{2}   H \eta_1\tau_2^{-1-\eta_1}e^{i \pi \eta_1-i k\tau_1} }{2 \sqrt{\epsilon_0}(1+\eta_1)(3+\eta_1) }(-\tau)^{3+\eta_1}
  	\right)k^{1/2}
  	\nonumber\\
  	&+&\left( \begin{split} &
  \frac{ c_{s_1}^2  H  \left((3+\eta_1)\tau_1^3-3\eta_1\tau_1\tau_2^2+2\eta_1\tau_2^3\right) e^{-i k\tau_1 }}{12 \sqrt{\epsilon_0}(3+\eta_1)}
 \\+&
  			\frac{  c_{s_1}^{2}  H  \left(\eta_1\tau_1 - \tau_2-\eta_1\tau_2\right)\tau_2^{-1-\eta_1}e^{i\pi \eta_1-i k \tau_1}}{2 \sqrt{\epsilon_0}(1+\eta_1)(3+\eta_1) }			
  		(-\tau)^{3+\eta_1}
	\end{split} \right) k^{3/2} \nonumber \\
   &+&\dots\;.
  \end{eqnarray}	
Clearly, the solution given in    Eq.~(\ref{B-R-IR}) contains a time-independent part and a time-dependent one, which will  grow with time when $\eta_1<-3$.

Using  $N_2$ and $N_3$ to denote the number of $e$-folds during the second and third phases, respectively,
i.e. $\tau_2=\tau_1 e^{-N_2}$ and $\tau=\tau_1 e^{-N_2-N_3}$, we obtain,  from Eq.~(\ref{B-R-IR}), the expression of the power spectrum of the curvature perturbations
	\begin{eqnarray}\label{B-P-IR}
		\dfrac{\mathcal{P}_{\widetilde{\mathcal{R}}^{(2)}_k}}{\mathcal{P}_0} &\simeq&
		1+\left(1-c_{s_1}^2+\frac{c_{s_1}^2\eta_1}{3+\eta_1} e^{-2N_2}+\frac{2c_{s_1}^2\eta_1}{(1+\eta_1)(3+\eta_1)} e^{-2N_2-\left(3+\eta_1 \right)N_3}\right)\tau_1^2k^2
		\nonumber\\	
				&-&
\left(\begin{split}&
			\frac{1}{3}+\frac{2\eta_1e^{-3N_2}}{9+3\eta_1}-\frac{3\eta_1e^{-2N_2}}{9+3\eta_1}
         +\frac{2+2\eta_1-2\eta_1 e^{N_2}}{(1+\eta_1)(3+\eta_1)} e^{-3N_2-\left(3+\eta_1\right)N_3}
			\\-&
			c_{s_1}^2\left(\frac{1}{2}-\frac{\eta_1}{6+2\eta_1}e^{-2N_2}-\frac{\eta_1}{(1+\eta_1)(3+\eta_1)} e^{-2N_2-\left(3+\eta_1\right)N_3} \right)^2
		\end{split}
			\right)c_{s_1}^2\tau_1^4k^4
		\nonumber\\
		&+&
		\left(
			\frac{1}{6}+\frac{\eta_1e^{-3N_2}}{9+3\eta_1}-\frac{\eta_1e^{-2N_2}}{6+2\eta_1}
         +\frac{1+\eta_1-\eta_1 e^{N_2}}{(1+\eta_1)(3+\eta_1)} e^{-3N_2-\left(3+\eta_1\right)N_3}
			\right)^2c_{s_1}^4\tau_1^6k^6
 	\;.
\end{eqnarray}

\begin{figure}[H]
 	\centering
 	\includegraphics[width=0.6\linewidth]{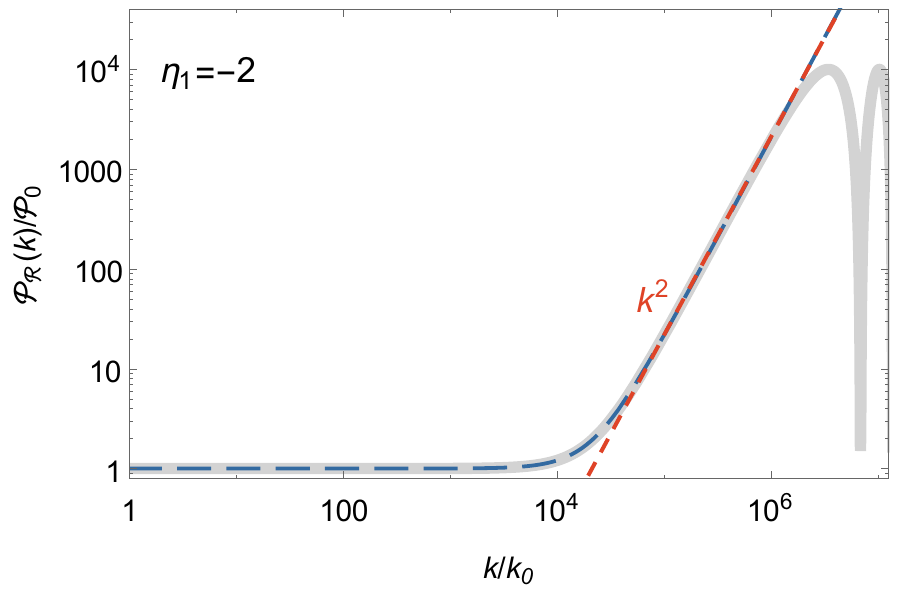}
 	\caption{\label{fig9}
The   power spectrum in the $\eta_1=-2$ case. The gray-solid and  blue-dashed lines represent the numerical  and approximate results, respectively. The red-dashed line indicates the $k^2$ growth.}
\end{figure}

We first study the $\eta_1>-3$ case, which means that there are no growing terms in the solution of the curvature perturbations and all time-dependent terms in Eq.~(\ref{B-R-IR}) decay with the cosmic expansion. Thus,  all terms containing $N_2$ and $N_3$ in Eq.~(\ref{B-P-IR}) can be neglected and the power spectrum can be simplified as
\begin{eqnarray}\label{B-P-1}
		\dfrac{\mathcal{P}_{\widetilde{\mathcal{R}}^{(2)}_k}}{\mathcal{P}_0} &\simeq&
		1+\tau_1^2k^2-\frac{1}{3}c_{s_1}^2\tau_1^4k^4+\frac{1}{36} c_{s_1}^4\tau_1^6k^6
 	\;.
\end{eqnarray}
It can be found that the wave numbers at which  the scale-invariant term is comparable to the $k^2$ term and the $k^2$ term is comparable to $k^4$ one happen, respectively, at
\begin{eqnarray}\label{}
k_{1}&\simeq& c_{s_1}k_c \;, \nonumber\\
k_{2}&\simeq & \sqrt{3} k_c\;.
\end{eqnarray}
The power spectrum has a growth rate of $k^2$ since $k_2>k_c$.
These results can be found clearly in Fig.~\ref{fig9},  where the evolutions of the power spectrum from numerical and approximate analyses are plotted in the $\eta_1=-2$ case.
At the CMB scale, the power spectrum is  scale invariant  which is consistent with the CMB observations. At scales smaller than the CMB scale, the power spectrum becomes scale-dependent with a $k^2$ growth.

When $\eta_1<-3$, since the coefficient of $k^n$ is very tedious,  we only consider  three special cases ($N_2\gg N_3$, $N_2\ll N_3$, and $N_2= N_3$) to investigate analytically the growth of the power spectrum.

First of all, when $N_2 \gg N_3$, all terms containing $N_2$ in Eq.~(\ref{B-P-IR}) can be neglected. Thus, the dominant term of $k^n$ is the same as the one in the case of  $\eta_1 > -3$. Therefore, the evolution of the power spectrum is  the same as in the case of $\eta_1 > -3$, and only has the $k^2$ growth.

\begin{figure}[H]
	\centering
	\includegraphics[width=0.3\linewidth]{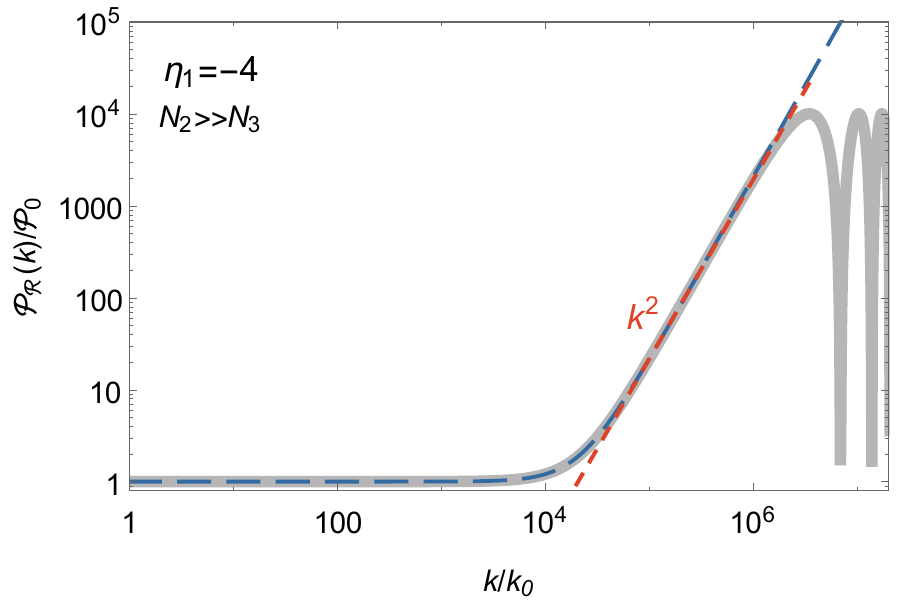}
 \includegraphics[width=0.3\linewidth]{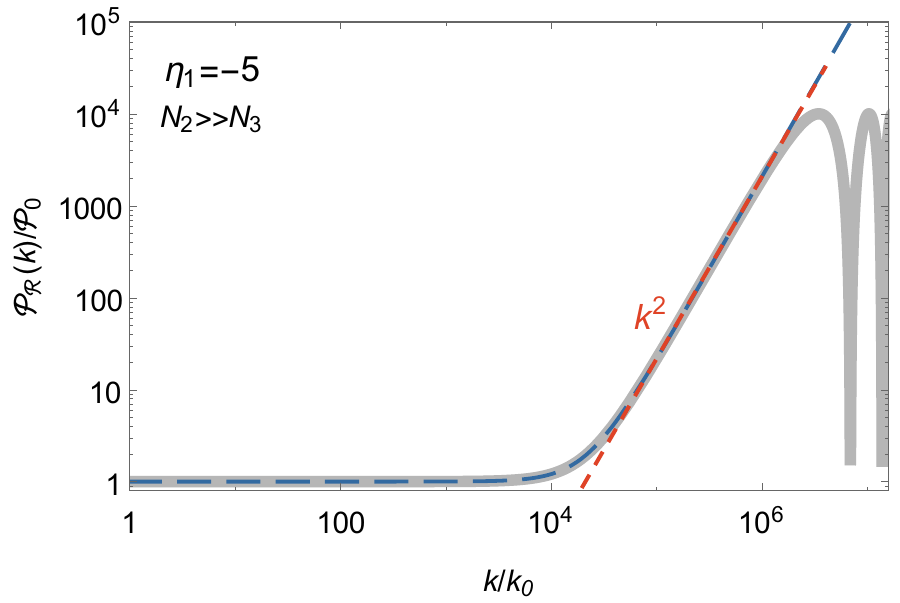}
 \includegraphics[width=0.3\linewidth]{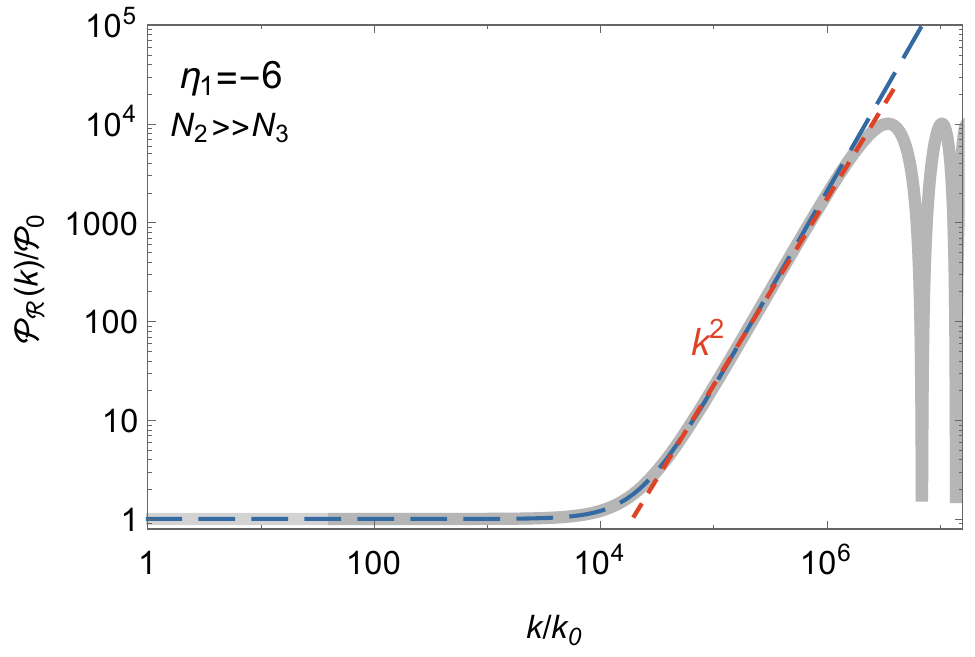}
	\includegraphics[width=0.3\linewidth]{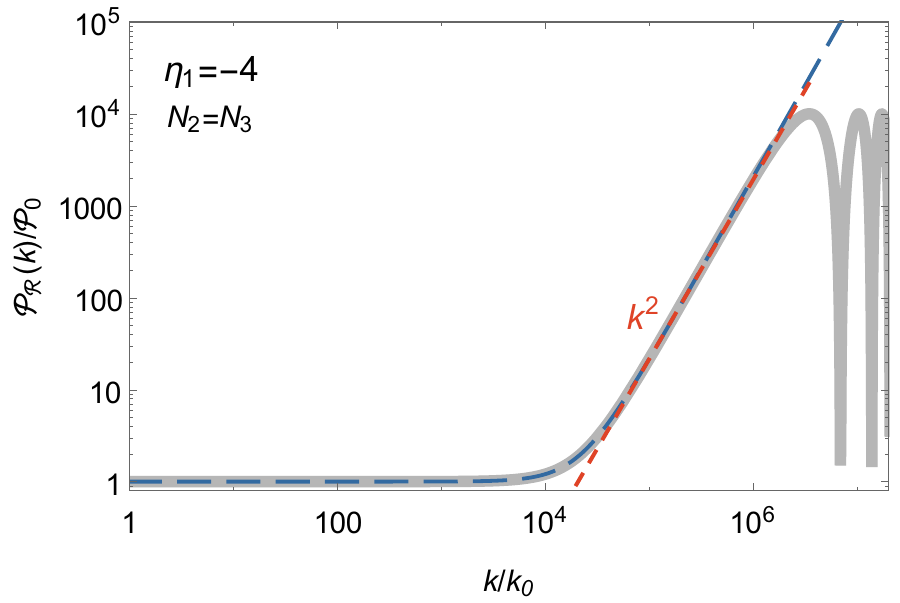}
	\includegraphics[width=0.3\linewidth]{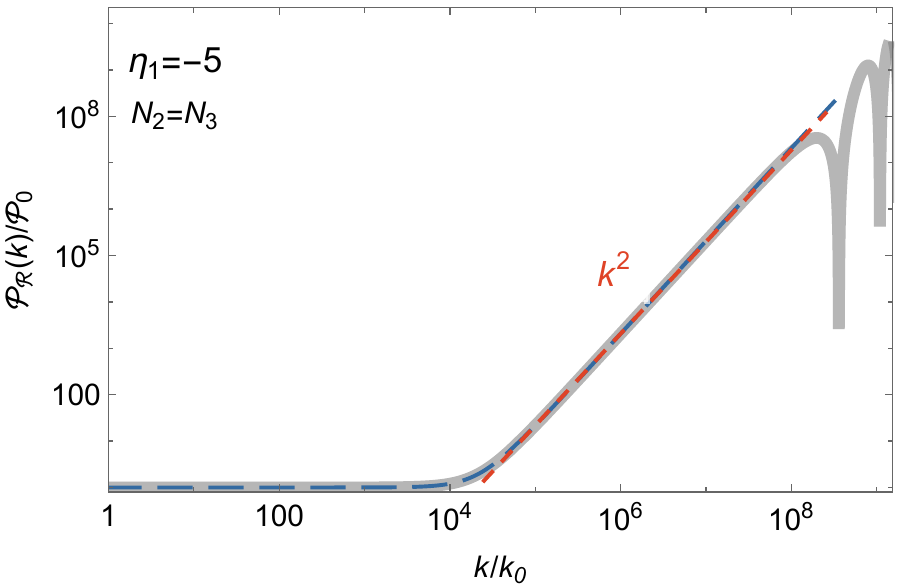}	
	\includegraphics[width=0.3\linewidth]{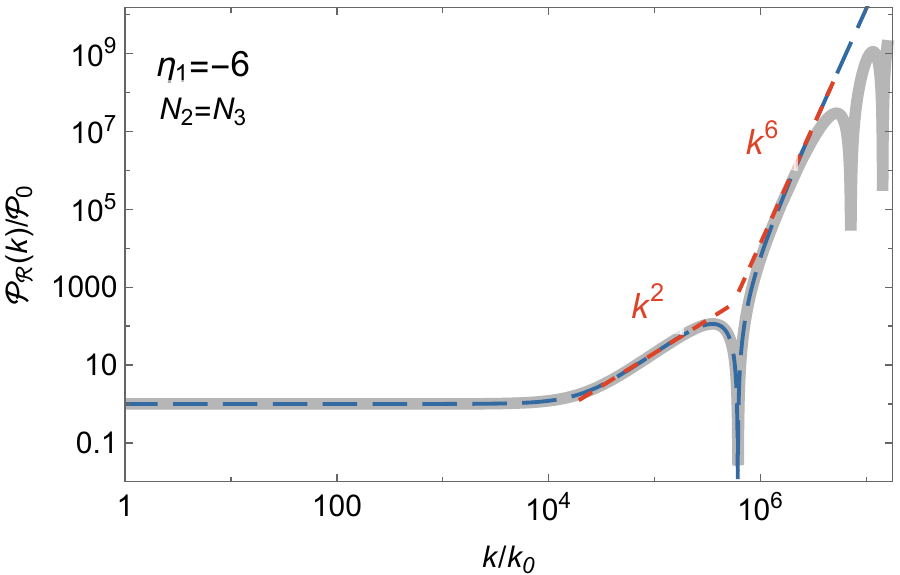}
	\caption{	\label{fig10}The  power spectrums  in the  $\eta_1=-4$, $-5$ and $-6$ cases. The gray-solid and  blue-dashed lines represent the numerical and approximate results, respectively.   	}
\end{figure}

When $N_2=N_3$, the expression for the power spectrum given in Eq.~(\ref{B-P-IR}) can be simplified to be 
\begin{eqnarray}\label{}
		\frac{\mathcal{P}_{\widetilde{\mathcal{R}}^{(2)}_k}}{\mathcal{P}_0} &\simeq&
		1+\tau_1^2k^2
		-
\left(
			\frac{1}{3}
         -\frac{2\eta_1}{(1+\eta_1)(3+\eta_1)} e^{-\left(5+\eta_1\right)N_3}
			\right)c_{s_1}^2\tau_1^4k^4
		\nonumber\\
		&+&
		\left(
			\frac{1}{36}
         -\frac{\eta_1 }{3(1+\eta_1)(3+\eta_1)} e^{-\left(5+\eta_1\right)N_3}
         +\frac{\eta_1^2 }{(1+\eta_1)^2(3+\eta_1)^2} e^{-2\left(5+\eta_1\right)N_3}
			\right)c_{s_1}^4\tau_1^6k^6
 	\;.
\end{eqnarray}
Since different values of $\eta_1$ will give different results,  
we will discuss this situation by considering following different cases:
	\begin{eqnarray}\label{}
		\frac{\mathcal{P}_{\widetilde{\mathcal{R}}^{(2)}_k}}{\mathcal{P}_0} \simeq \left\{\begin{array}{ll}
1+\tau_1^2k^2-\dfrac{1}{3}c_{s_1}^2\tau_1^4k^4+\dfrac{1}{36} c_{s_1}^4\tau_1^6k^6
& -5<\eta_1<-3
\vspace{0.4cm} \\
			1+\tau_1^2k^2-\dfrac{19}{12}c_{s_1}^2\tau_1^4k^4+\dfrac{361}{576} c_{s_1}^4\tau_1^6k^6
			&~~~~ \eta_1=-5	
			\vspace{0.4cm} \\
			1+\tau_1^2k^2
		+
   \dfrac{2\eta_1 e^{-\left(5+\eta_1\right)N_3}}{(1+\eta_1)(3+\eta_1)}
			c_{s_1}^2\tau_1^4k^4
		+	
         \dfrac{\eta_1^2 e^{-2\left(5+\eta_1\right)N_3}}{(1+\eta_1)^2(3+\eta_1)^2}
			c_{s_1}^4\tau_1^6k^6
			&~~~~ \eta_1<-5\ .	 \nonumber\\				
			\end{array} \right.
	\end{eqnarray}
When $-5\leq \eta_1<-3$, we find that $k_2 \simeq k_c$, which means that the  power spectrum only has the $k^2$ growth.  
In the $\eta_1<-5$ case, we obtain
\begin{eqnarray}\label{k_62}
	k_{1}  &\simeq&  c_{s_1} k_c \;,\nonumber\\
    k_{2}  &\simeq&\frac{\sqrt{(1+\eta_1)(3+\eta_1)}}{ \sqrt{-2\eta_1} } e^{\frac{1}{2}(5+\eta_1)N_3} k_c \;,\nonumber\\
    k_{3}  &\simeq&  \frac{\sqrt{2(1+\eta_1)(3+\eta_1)}}{ \sqrt{-\eta_1}} e^{\frac{1}{2}(5+\eta_1)N_3}k_c\;.
\end{eqnarray}
Apparently,  $k_3$ is less than  $ k_c$ if $ \frac{\sqrt{2(1+\eta_1)(3+\eta_1)}}{ \sqrt{-\eta_1}} e^{\frac{1}{2}(5+\eta_1)N_3 }\ll 1$, which can be realized easily since $\eta_1<-5$. Thus, the power spectrum can have the $k^6$ growth.
 Since there is a negative coefficient in the $k^4$ term, the power spectrum has a dip preceding the $k^6$ growth. 
In Fig.~\ref{fig10}, the power spectrums in the cases of $\eta_1=-4,-5$ and $-6$ for the $N_2\gg N_3$ and  $N_2=N_3$   cases are plotted.
We can see that when $N_2 \gg N_3$, the highest growth slope of the power spectrum is  $k^2$. In the case of $N_2=N_3$, the highest growth rate for $\eta_1=-4$ and $-5$ is still $k^2$, while for $\eta_1=-6$, the highest growth rate can reach up to $k^6$.

\begin{figure}[H]
	\centering
	\includegraphics[width=0.3\linewidth]{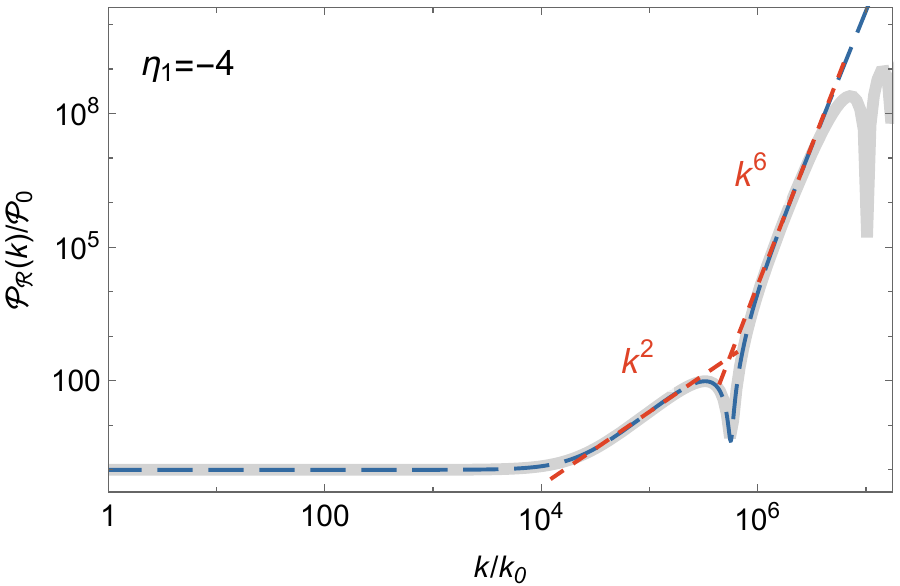}
	\includegraphics[width=0.3\linewidth]{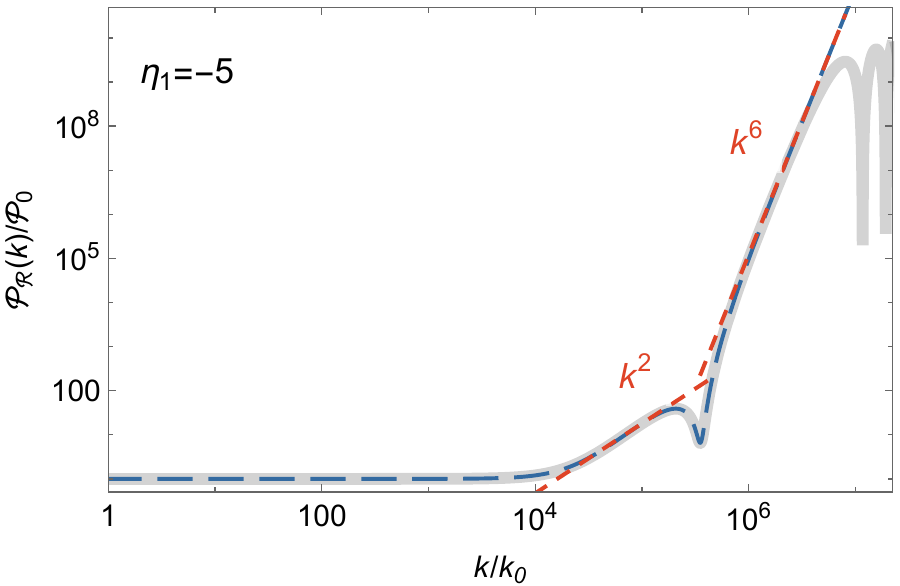}
	\includegraphics[width=0.3\linewidth]{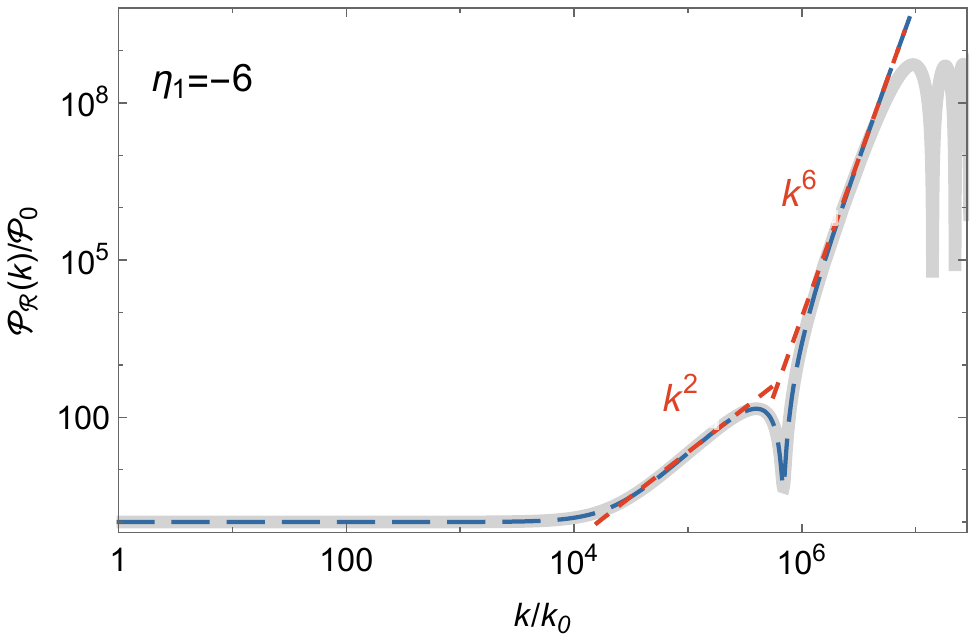}
\includegraphics[width=0.3\linewidth]{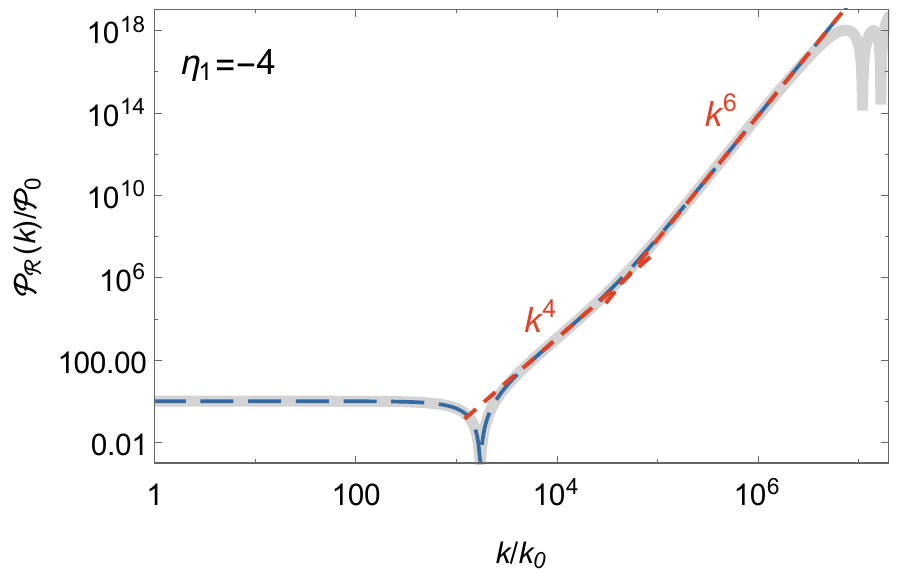}
\includegraphics[width=0.3\linewidth]{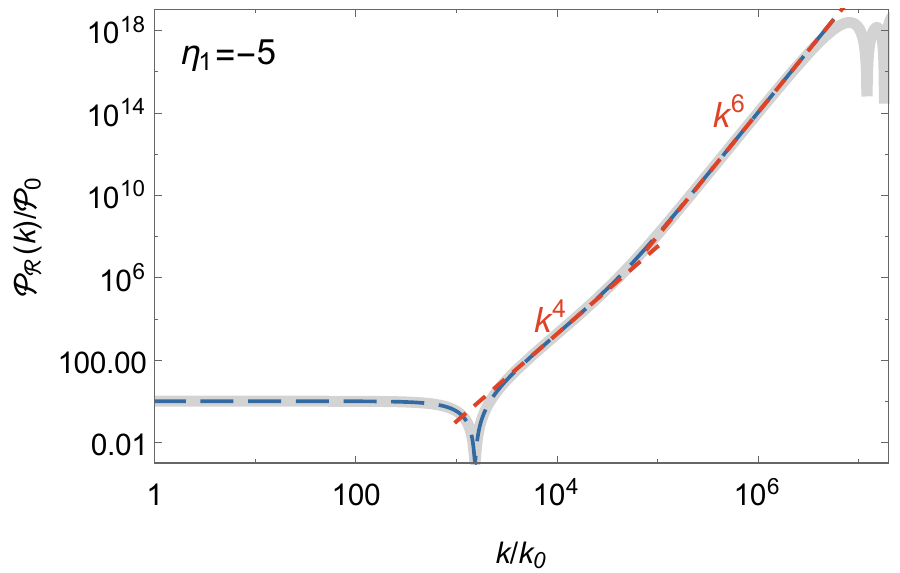}
    \includegraphics[width=0.3\linewidth]{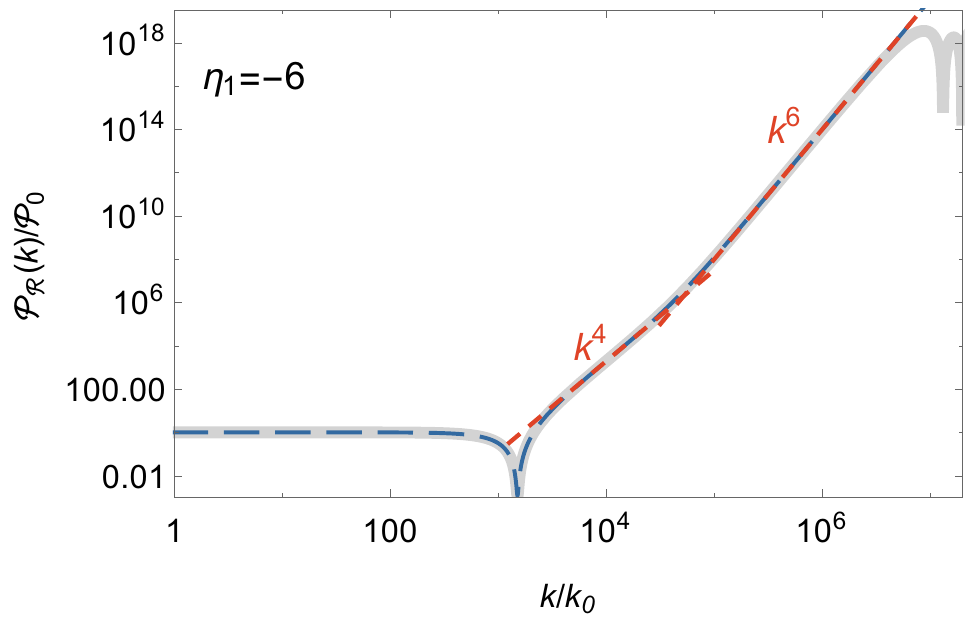}
	\caption{	\label{fig11}
The   power spectrums for $N_2\ll N_3$ in the  $\eta_1=-4,-5$ and $-6$ cases. The gray-solid and  blue-dashed lines represent the numerical and approximate  results, respectively.
The first line is corresponding to the $c_{s_1}^2 e^{-(3+\eta_1)N_3}\ll 1 $ case and the second line to the $c_{s_1}^2 e^{-(3+\eta_1)N_3}\gg 1 $ case.
	}
\end{figure}

When  $N_ 2\ll N_ 3 $, the power spectrum can be approximated as
	\begin{eqnarray}\label{}
		\frac{\mathcal{P}_{\widetilde{\mathcal{R}}^{(2)}_k}}{\mathcal{P}_0} &\simeq&
		1+\left(1+\frac{2c_{s_1}^2\eta_1}{(1+\eta_1)(3+\eta_1)} e^{-\left(3+\eta_1 \right)N_3}\right)\tau_1^2k^2
		\nonumber\\	
				&+&
\left(
         \frac{2\eta_1}{(1+\eta_1)(3+\eta_1)} e^{-\left(3+\eta_1\right)N_3}
			+
			\frac{c_{s_1}^2\eta_1^2}{(1+\eta_1)^2(3+\eta_1)^2} e^{-2\left(3+\eta_1\right)N_3}
			\right)c_{s_1}^2\tau_1^4k^4
		\nonumber\\
		&+&
         \frac{\eta_1^2 }{(1+\eta_1)^2(3+\eta_1)^2} e^{-2\left(3+\eta_1\right)N_3}
			c_{s_1}^4\tau_1^6k^6
 	\;.
\end{eqnarray}
If $c_{s_1}^2 e^{-(3+\eta_1)N_3}\ll 1 $, the wave numbers when the $k^n$ and $k^{n-2}$ terms become comparable happen at
\begin{eqnarray}\label{k_62}
	k_{1}  &\simeq & c_{s_1}k_c \;,\nonumber\\
    k_{2}  &\simeq& \frac{\sqrt{(1+\eta_1)(3+\eta_1)}}{ \sqrt{-2\eta_1} } e^{\frac{1}{2}(3+\eta_1)N_3} k_c \;,\nonumber\\
    k_{3}  &\simeq&  \frac{\sqrt{2(1+\eta_1)(3+\eta_1)}}{ \sqrt{-\eta_1}} e^{\frac{1}{2}(3+\eta_1)N_3}k_c\;.
\end{eqnarray}
Apparently,  $k_3 < k_c$ can be satisfied easily since $\eta_1<-3$  and thus  the power spectrum can have  a $k^6$ growth. The dip will appear preceding the $k^6$ growth since the $k^4$ term has a negative coefficient. 
If $c_{s_1}^2 e^{-(3+\eta_1)N_3}\gg 1 $, we obtain
\begin{eqnarray}\label{k_62}
	k_{1}  &\simeq&\frac{\sqrt{(1+\eta_1)(3+\eta_1)}}{ \sqrt{-2\eta_1} } e^{\frac{1}{2}(3+\eta_1)N_3} k_c\;,\nonumber\\
    k_{2}  &\simeq& \frac{\sqrt{2(1+\eta_1)(3+\eta_1)}}{ \sqrt{-\eta_1} } e^{\frac{1}{2}(3+\eta_1)N_3} k_c\;,\nonumber\\
    k_{3}  &\simeq& c_{s_1}k_c\;.
\end{eqnarray}
It can be found that $k_3<k_c$ and $k_1\approx k_2$, which means that the power spectrum will enter $k^4$ growth directly after the scale-invariant spectrum, and will enter finally the $k^6$ growth.  The power spectrum has the dip preceding the $k^4$ growth due to the negative coefficient in the $k^2$ term. 
These characters can be found in Fig.~\ref{fig11}, where the power spectrums in the cases of $\eta_1=-4,-5$ and $-6$ for the $N_2\ll N_3$ are plotted. 

\begin{figure}[H]
 	\centering
 	\includegraphics[width=0.4\linewidth]{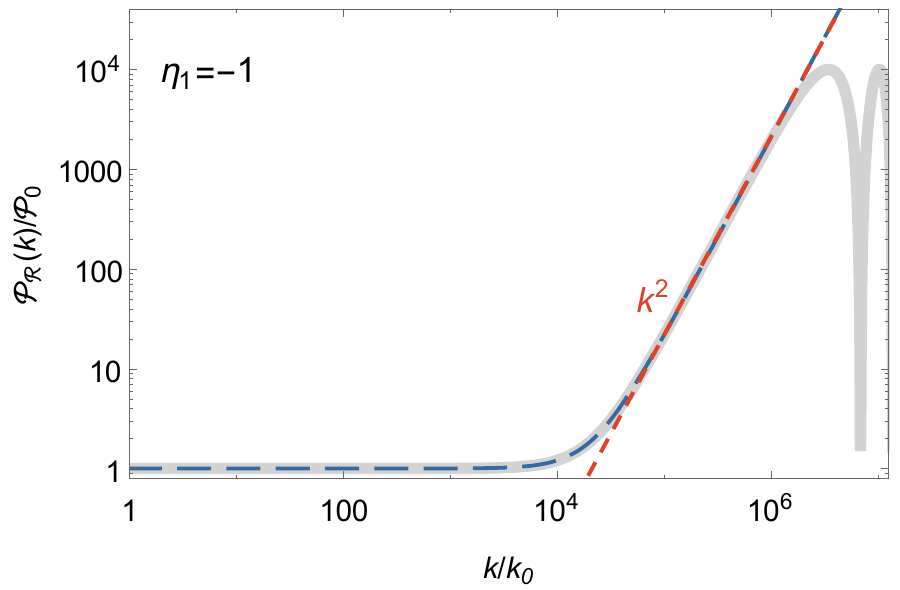}
\includegraphics[width=0.4\linewidth]{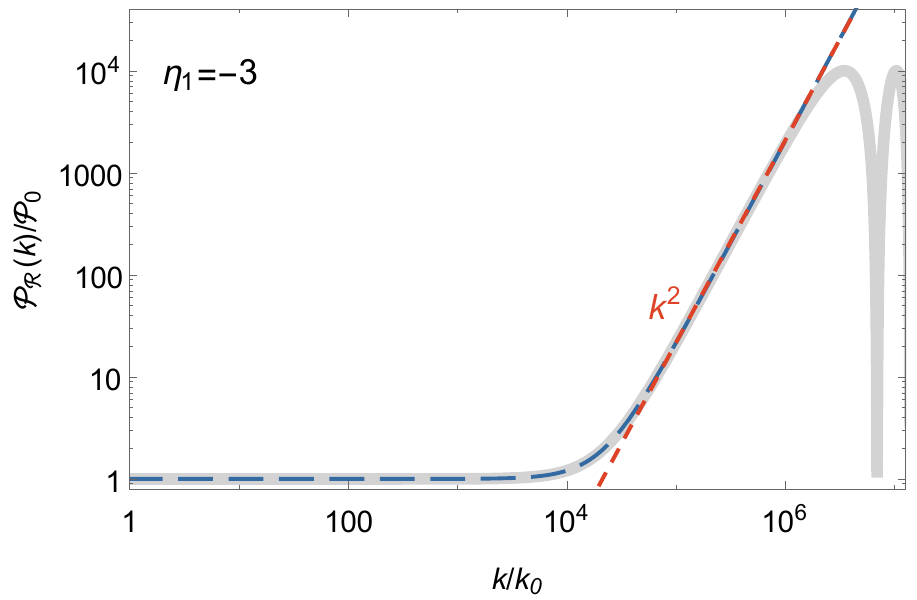}
 	\caption{\label{fig12}
 			The   power spectrums  in the  $\eta_1=-1$ and $\eta_1=-3$ cases. The gray-solid and  blue-dashed lines represent the numerical and  approximate  results, respectively. 		}
 \end{figure}

Furthermore,  when $\eta_1=-1$ and $-3$ there are singularities in Eq.~(\ref{B-R-IR}) and thus these cases  need to be studied separately.
To avoid the problem, we  first set the value of $\eta_1$, and then expand  Eq.~(\ref{RB3-R}) in the infrared limit  to obtain
{\small
	\begin{eqnarray}\label{RB3-et1}
		\widetilde{\mathcal{R}}^{(2)}_k (\tau)  = \left\{\begin{array}{ll}
			\dfrac{i He^{-i k \tau_1}}{2 \sqrt{\epsilon_0 k^3}}-\dfrac{ H\tau_1 e^{-i k\tau_1 }}{2  \sqrt{\epsilon_0 k}}-
			\dfrac{iHc_{s_1}^2 e^{-i k\tau_1 }k^{1/2}}{8 \sqrt{\epsilon_0}}
			\left[2\tau_1^2+\tau_2^2-(-\tau)^2\right]
			+\dots
			& \eta_1=-1	
			\vspace{0.4cm} \\
			\dfrac{i He^{-i k \tau_1}}{2 \sqrt{\epsilon_0 k^3}}-\dfrac{ H\tau_1 e^{-i k\tau_1 }}{2  \sqrt{\epsilon_0 k}}-
			\dfrac{iHc_{s_1}^2 e^{-i k\tau_1 }k^{1/2}}{8 \sqrt{\epsilon_0}}
			\left[2\tau_1^2-3\tau_2^2-6\tau_2^2\log \left(\dfrac{\tau}{\tau_2}\right)\right]
			+\dots
			& \eta_1=-3\ .	 \nonumber\\				
			\end{array} \right.
	\end{eqnarray}
}
In the case of $\eta_1=-1$, the solution consists of   constant and  decaying terms, and thus it is dominated by the constant terms. When $\eta_1=-3$, except for the constant terms, the solution of  the curvature perturbations contains  a logarithmic-growing term. This character is different from that in the $\eta_1<-3$ case, where the growth of the  solution is  power law. 
Since the coefficient of the logarithmic-growing term depends on $c_{s_1}^2$, which  is much less than one in our analysis, the contribution of the growth term in the solution is negligible. Thus, from Eq.~(\ref{RB3-et1}), we obtain that the power spectrum has the same expression  
	\begin{eqnarray}\label{RB3-et1}
		\dfrac{\mathcal{P}_{\widetilde{\mathcal{R}}^{(2)}_{k}}}{\mathcal{P}_0} \simeq  
			1+\tau_1^2k^2+\dfrac{1}{4}c_{s_1}^4\tau_1^4 k^4 
	\end{eqnarray}
for  $\eta_1=-1$ and $-3$. The steepest growth is $k^2$ apparently.  The corresponding numerical and approximate  results of the power spectrum are shown  in Fig.~\ref{fig12}.   

\subsection{Changes of slow-roll parameter $\eta$ previous to the sound speed}
\label{sec-eta-cs}

\begin{figure}[H]
	\centering
	\includegraphics[width=0.6\linewidth]{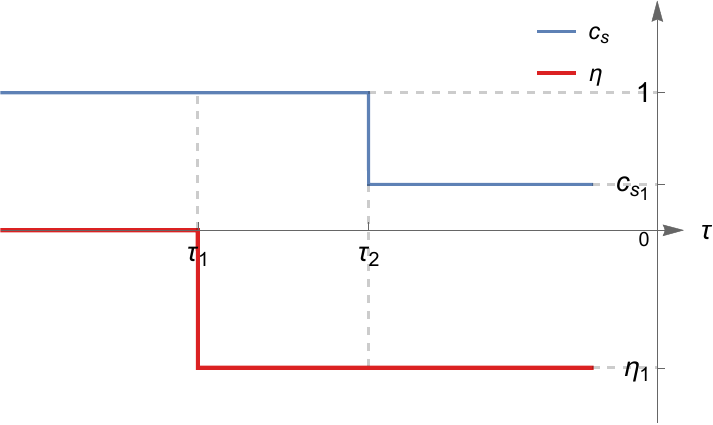}
	\caption{\label{fig13}
		  The sound speed $c_s$ (blue line)   varies  at $\tau_2$, and the slow-roll parameter $\eta$ (red line) changes at $\tau_1$   $(|\tau_2|<|\tau_1|$).
	}
\end{figure}

This scenario is shown in Fig.~\ref{fig13}. The slow-roll parameter $\eta$ changes  from $\sim 0$ to a negative constant $\eta_1$ at $\tau_1$, and the sound speed decreases  from $1$ to a  small constant $c_{s_1}$ at $\tau_2$ ($|\tau_2|<|\tau_1|$).

\begin{figure}[H]
	\centering
	\includegraphics[width=0.4\linewidth]{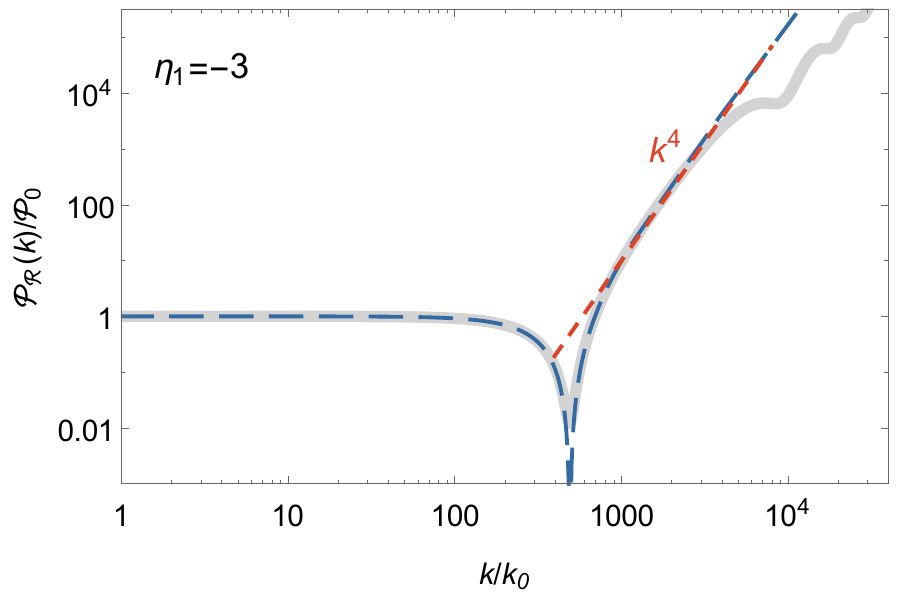}
	\includegraphics[width=0.4\linewidth]{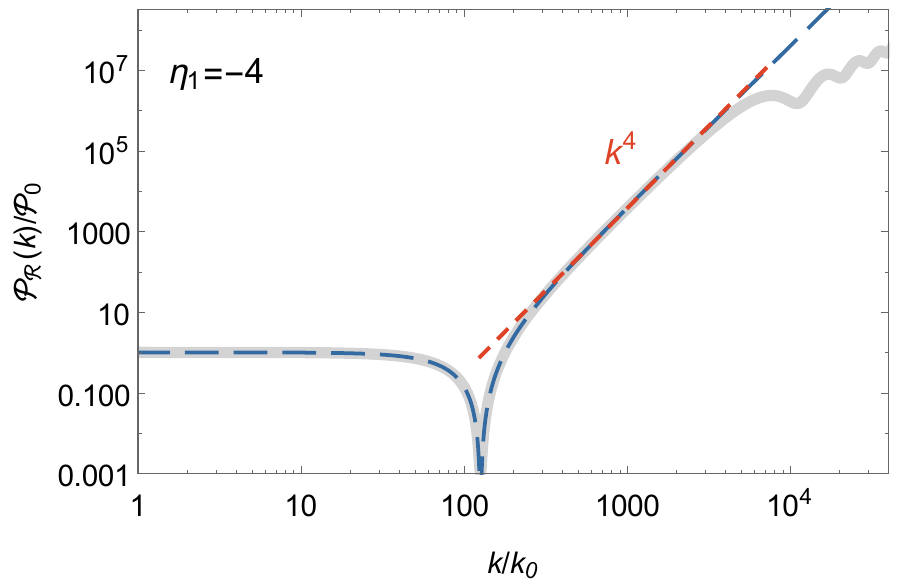}
	\includegraphics[width=0.4\linewidth]{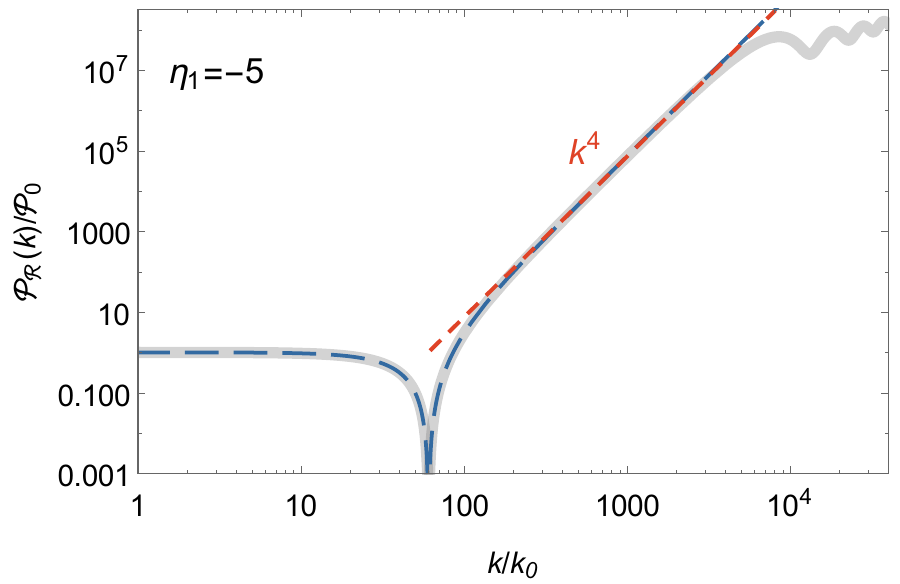}
	\includegraphics[width=0.4\linewidth]{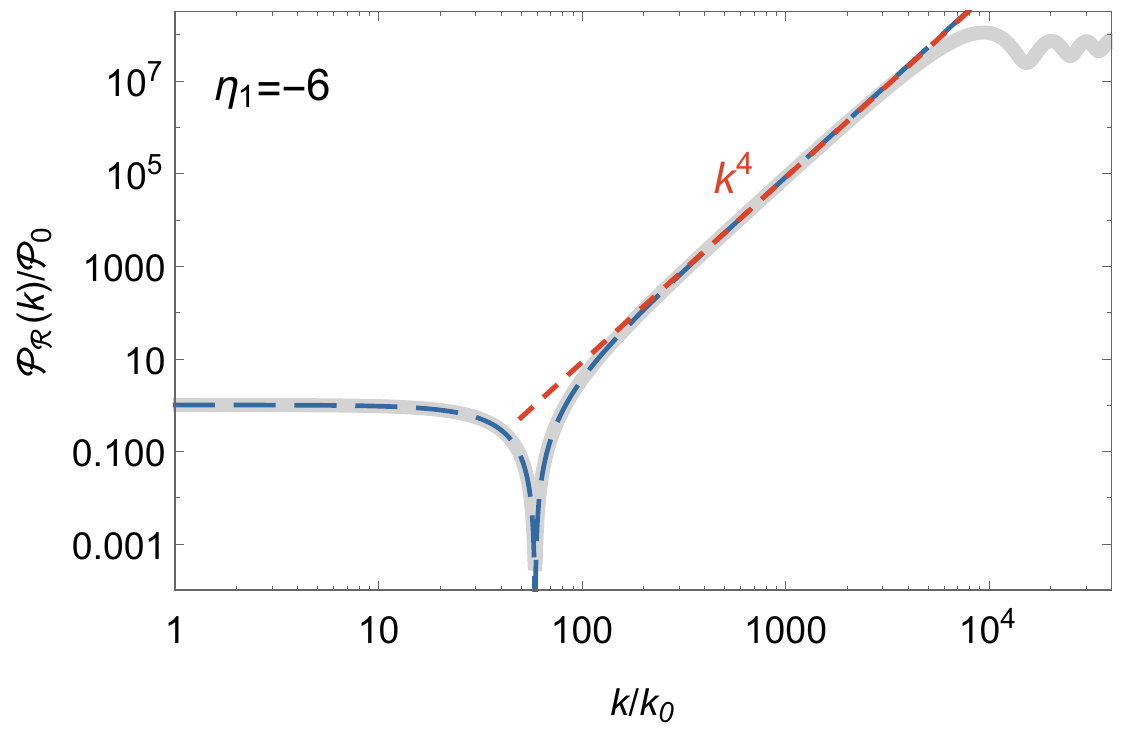}
	\caption{\label{fig14}  	
		The  power spectrums   for different constant values of $\eta_1$. The gray-solid and  blue-dashed lines represent the numerical  and approximate results, respectively.
	}
\end{figure}

Since the method for analytical solution in this case is similar to that of the preceding subsection,  we do not  give the details here. The general expression of the curvature perturbations  in the infrared region ($-c_{s_1} k \tau \rightarrow 0$, $ -c_{s_1} k \tau_2 \rightarrow 0$ and $ -k \tau_1 \rightarrow 0$) is so complex,  and we do not show it here. We only consider the integer $\eta_1$ cases.
Different from the results obtained in the preceding subsection,  we find that the analytical results  do not coincide with the numerical ones when $\eta_1=-1$ and $-2$.
Therefore, we only give the infrared expressions ($-k\tau_1\rightarrow 0, -c_{s_1} k \tau_2\rightarrow 0$ and $-c_{s_1} k \tau\rightarrow 0$) of  the curvature perturbations in the  $\eta_1\leq-3$ case:
{
\begin{eqnarray}\label{RC3}
		\hat{\mathcal{R}}^{(2)}_k (\tau)  = \left\{\begin{array}{ll}
			\dfrac{i  H}{2 \sqrt{\epsilon_0 k^3}}
\\			+
			\left(\dfrac{i H \left(3\tau_1^2-\left(1-c_{s_1}^2\right)\tau_2^2 \right)}{8 \sqrt{\epsilon_0}}+\dfrac{3 i H \tau_1^2}{ 4\sqrt{\epsilon_0}}\left(c_{s_1}^2 \log \left[\dfrac{\tau}{\tau_2}\right] -
			 \log \left[\dfrac{\tau_1}{\tau_2}\right]
			\right)\right)k^{1/2}\\
			+\left(\dfrac{ H \tau_1^3}{6 \sqrt{\epsilon_0}}-\dfrac{  H \tau_1^3}{ 2\sqrt{\epsilon_0}}\log \left[\dfrac{\tau_1}{\tau_2}\right]+\dfrac{ c_{s_1}^2 H \tau_1^3}{ 2\sqrt{\epsilon_0}}\log \left[\dfrac{\tau}{\tau_2}\right]
		\right)k^{3/2}+\dots
			& \eta_1=-3
			\vspace{0.4cm} \\
			\dfrac{i  H}{2 \sqrt{\epsilon_0 k^3}}
			\\-
			\left(\dfrac{i H \left(\left(1-c_{s_1}^2\right)\left(8\tau_1^3 +\tau_2^3\right)-12\tau_1^2\tau_2 \right)}{12 \sqrt{\epsilon_0}\tau_2}-\dfrac{2ic_{s_1}^2 H \tau_1^3}{ 3\sqrt{\epsilon_0}}(-\tau)^{-1}\right)k^{1/2}. \\
			-\left(\dfrac{ H \tau_1^3 \left(3\left(1-c_{s_1}^2\right)\tau_1-4\tau_2 \right)}{6 \sqrt{\epsilon_0}\tau_2}-\dfrac{ c_{s_1}^2 H \tau_1^4}{ 2\sqrt{\epsilon_0}}\left(-\tau\right)^{-1}
			\right)k^{3/2}+\dots
			& \eta_1=-4
			\vspace{0.4cm} \nonumber \\
			\dfrac{i  H}{2 \sqrt{\epsilon_0 k^3}}
			\\-
			\left(\dfrac{i H \left(\left(1-c_{s_1}^2\right)\left(5\tau_1^4 +\tau_2^4\right)-10\tau_1^2\tau_2^2 \right)}{16 \sqrt{\epsilon_0}\tau_2^2}+\dfrac{5 i c_{s_1}^2 H \tau_1^4}{ 16\sqrt{\epsilon_0}}(-\tau)^{-2}\right)k^{1/2}\\
			-\left(\dfrac{ H \tau_1^3 \left(3\left(1-c_{s_1}^2\right)\tau_1^2-5\tau_2^2 \right)}{12 \sqrt{\epsilon_0}\tau_2^2}+\dfrac{ c_{s_1}^2 H \tau_1^5}{ 4\sqrt{\epsilon_0}}\left(-\tau\right)^{-2}
			\right)k^{3/2}+\dots
			&\eta_1=-5
			\vspace{0.4cm} \\
			\dfrac{i  H}{2 \sqrt{\epsilon_0 k^3}}
			\\-
			\left(\dfrac{i H \left(\left(1-c_{s_1}^2\right)\left(4\tau_1^5 +\tau_2^5\right)-10\tau_1^2\tau_2^3 \right)}{20 \sqrt{\epsilon_0}\tau_2^3}-\dfrac{5 i c_{s_1}^2 H \tau_1^5}{ 5\sqrt{\epsilon_0}}(-\tau)^{-3}\right)k^{1/2}\\
			-\left(\dfrac{ H \tau_1^3 \left(\left(1-c_{s_1}^2\right)\tau_1^3-2\tau_2^3 \right)}{6 \sqrt{\epsilon_0}\tau_2^3}-\dfrac{ c_{s_1}^2 H \tau_1^6}{ 6\sqrt{\epsilon_0}}\left(-\tau\right)^{-3}
			\right)k^{3/2}+\dots
			& \eta_1=-6\ \;.
		\end{array} \right.
	\end{eqnarray}
}
From the above expression,  one can  obtain the power spectrum:
	\begin{eqnarray}\label{PRC3}
	\frac{\mathcal{P}_{	\hat{\mathcal{R}}^{(2)}_k}}{\mathcal{P}_0}  \simeq \left\{\begin{array}{ll}
		1+\frac{1}{2}\left(3-e^{-2N_2}-6N_2-6c_{s_1}^2 N_3\right)\tau_1^2k^2
	 \\
		~+
		\frac{1}{16}\left(3-e^{-2N_2}-6N_2-6c_{s_1}^2 N_3\right)^2\tau_1^4k^4 \\
		~+\frac{1}{9}\left(1-3N_2-3c_{s_1}^2N_3\right)^2\tau_1^6k^6
			& \eta_1=-3
			\vspace{0.4cm} \\
		1+\frac{1}{3}\left(12-e^{-2N_2} -8e^{N_2}+8c_{s_1}^2 e^{N_2} -8c_{s_1}^2 e^{N_2+N_3}\right)\tau_1^2k^2
		 \\
		~+
		\frac{1}{36}\left(12-e^{-2N_2} -8e^{N_2}+8c_{s_1}^2 e^{N_2} -8c_{s_1}^2 e^{N_2+N_3}\right)^2\tau_1^4k^4
		 \\
		~+
		\frac{1}{9}\left(4-3e^{N_2}+3c_{s_1}^2 e^{N_2} -3c_{s_1}^2 e^{N_2+N_3}\right)^2\tau_1^6k^6
			& \eta_1=-4
			\vspace{0.4cm} \\
			1+\frac{1}{4}\left(10-e^{-2N_2} -5e^{2N_2}+5c_{s_1}^2 e^{2N_2} -5c_{s_1}^2 e^{2N_2+2N_3}\right)\tau_1^2k^2
		 \\
		~+
		\frac{1}{64}\left(10-e^{-2N_2} -5e^{2N_2}+5c_{s_1}^2 e^{2N_2} -5c_{s_1}^2 e^{2N_2+2N_3}\right)^2\tau_1^4k^4
		 \\
		~+
		\frac{1}{36}\left(5-3e^{2N_2}+3c_{s_1}^2 e^{2N_2} -3c_{s_1}^2 e^{2N_2+2N_3}\right)^2\tau_1^6k^6
			&\eta_1=-5
			\vspace{0.4cm} \\
				1+\dfrac{1}{5}\left(10-e^{-2N_2} -4e^{3N_2}+4c_{s_1}^2 e^{3N_2} -4c_{s_1}^2 e^{3N_2+3N_3}\right)\tau_1^2k^2
		 \\
		~+
		\frac{1}{100}\left(10-e^{-2N_2} -4e^{3N_2}+4c_{s_1}^2 e^{3N_2} -4c_{s_1}^2 e^{3N_2+3N_3}\right)^2\tau_1^4k^4
		 \\
		~+
		\frac{1}{9}\left(2-e^{3N_2}+c_{s_1}^2 e^{3N_2} -c_{s_1}^2 e^{3N_2+3N_3}\right)^2\tau_1^6k^6
			& \eta_1=-6\ \;.
		\end{array} \right.
	\end{eqnarray}
Here $N_2$ and $N_3$ are the number of $e$-folds during the second and third phase, respectively.
The maximum wave number in the infrared  limit is $k |\tau_1|$. So, in Eq.~(\ref{PRC3}), the wave number must satisfy $k\ll \bar{k}_c\equiv -1/\tau_1$.
When the $k^6$ dominant term becomes comparable to the $k^4$ one, the wave number should be  equal to about
\begin{eqnarray}\label{k_6}
	k_{3}  \simeq \left\{\begin{array}{ll}
		-\dfrac{3}{2\tau_1}=
		\dfrac{3}{2}\bar{k}_c ~~~~& \eta_1=-3
		\vspace{0.2cm} \\
		-\dfrac{4}{3 \tau_1}=\dfrac{4}{3 }\bar{k}_c &\eta_1=-4
		\vspace{0.2cm} \\
		-\dfrac{5}{4 \tau_1}=\dfrac{5}{4 }\bar{k}_c &\eta_1=-5
		\vspace{0.2cm} \\
		-\dfrac{6}{5 \tau_1}=\dfrac{6}{5 }\bar{k}_c &\eta_1=-6\ .
	\end{array} \right.
\end{eqnarray}
It is obvious that all wave number  $k_3$ are  larger than $\bar{k}_c$ and thus are beyond the  infrared region, which means that  the steepest  growth of  the power spectrum is $k^4$.
Furthermore, we find that the power spectrum will  dip before the $k^4$ growth. 
The corresponding numerical and approximate results are shown in Fig.~\ref{fig14}.

 \section{conclusions}
\label{conclusion}
The generation of a significant amount of primordial black holes requires a sufficiently large power spectrum of the curvature perturbations of the order of about $\mathcal{O} (10^{-2})$ at the scales smaller than the CMB one.
There are two natural ways to amplify  the curvature perturbations. One is to reduce the rolling speed of the inflaton  and the other to suppress the sound speed $c_s$ of the curvature perturbations.
In the ultraslow-roll inflation scenario, it has been found that the power spectrum of the curvature perturbations has the $k^4$ growth.
In this paper, we use the improved junction conditions  to find that the power spectrum of the curvature perturbation has a $k^2$ growth when the speed of sound decreases suddenly.
Furthermore, by investigating the evolution of  the power spectrum in the inflation model, which can realize  decrease of both the sound speed and the rolling speed of the inflaton,
we find that the power spectrum at the large scales is nearly scale invariant to satisfy the constraint from the CMB observations, and at the same time it will be enhanced at the small scales to achieve an abundant formation of primordial black holes.
In the cases that the change of the slow-roll parameter $\eta$ precedes that of the sound speed $c_s$, the power spectrum of the curvature perturbations only has a $k^4$ growth. While if
 $\eta$ and  $c_s$ changes simultaneously or
the change of $c_s$  precedes that of $\eta$,  the power spectrum can possess   a $k^6$ growth under certain conditions, which is the steepest growth of the power spectrum reported analytically so far. In the multifield inflation, numerical results show that the growth larger than $k^6$ is possible due to the violent tachyonic instability~\cite{Braglia2020, Palma2020,Fumagalli2023,Braglia2021,Braglia2023}.

 \begin{acknowledgments}
We appreciate very much the insightful comments and helpful suggestions by the anonymous referee. This work is supported by the National Key Research and Development Program of China Grant No. 2020YFC2201502, and by the National Natural Science Foundation of China under Grants No. 12275080 and No.  12075084. 
\end{acknowledgments}

\appendix



\end{document}